# SPECIAL RELATIVITY DERIVED FROM CELLULAR AUTOMATA THEORY:
## The origin of the universal speed limit

**Tom Ostoma and Mike Trushyk**

48 O'HARA PLACE, Brampton, Ontario, L6Y 3R8
emqg@rogerswave.ca

Friday, January 22, 1999

### ACKNOWLEDGMENTS

We wish to thank R. Mongrain (P.Eng) for our lengthy conversations on the nature of space, time, light, matter, and CA theory. Also, special thanks goes to L. Walker for his help in proof reading this document.

# 1. ABSTRACT

*A new formulation of special relativity is described. It is based on a postulate that the universe is a vast Cellular Automata (CA), (ref. 2,3,4). It is also based on a new theory of inertia (ref. 5) proposed by R. Haisch, A. Rueda, and H. Puthoff, which we modified, and called Quantum Inertia (QI). In our theory of Quantum Inertia, we found that Newtonian Inertia is due to the strictly local electromagnetic force interactions of matter (quantum particles) with the surrounding charged virtual (matter) particles of the quantum vacuum. The sum of all the tiny electromagnetic forces originating from each charged particle in the mass with respect to the vacuum, is the source of the <u>total inertial force</u> of the mass which opposes accelerated motion in Newton's law 'F = MA'. Thus, QI resolves the problems and paradoxes of accelerated motion introduced in Mach's principle by suggesting that the state of acceleration of the charged virtual particles of the quantum vacuum with respect to the mass serves as Newton's universal reference frame (which Newton called 'absolute' space). The (net statistical) acceleration of the (charged) virtual particles of the quantum vacuum can be used as an absolute reference frame to gauge inertial mass. Therefore this frame can be used to define both absolute acceleration and <u>absolute mass,</u> which is equivalent to the relativistic rest mass. However, this frame <u>cannot</u> be used to gauge the absolute state of motion of an inertial reference frame. Thus Einstein's principle of relativity is still applicable for inertial frames (frames of constant velocity motion, or where Newton's law of inertia applies). The special relativistic treatment of inertial force, acceleration, and inertial mass is revised here to acknowledge the existence of absolute mass. We found that the special relativistic variation of mass with relative velocity $m = m_0 (1 - v^2/c^2)^{-1/2}$ is actually caused by <u>the decrease in the effectiveness of the applied force,</u> where the applied force and destination mass have a large relative velocity 'v'. We show that this decrease of the applied force in the reference frame of the moving mass is caused by a relativistic timing effect of the received force exchange particles, which alters the received flux rate of the force exchange bosons.*

*Minkowski 4D space-time is a simple consequence of the behavior of matter (elementary particles) and light (photons) on the Cellular Automata. At the tiniest quantum distance scales there exists a kind of secondary (quantized) absolute 3D space and separate absolute (quantized) time, as required by CA theory. This is represented by a rectangular 3D array of numbers or cells; $C(x,y,z)$. These cells change state after every new CA clock operation $\Delta t$. The array of numbers $C(x,y,z)$ is called CA space, which acts like the Newtonian version of Cartesian absolute space. There also exists a separate absolute time needed to evolve the numerical state of the CA. 3D CA space and time are not effected by any physical interactions, and are also not accessible through direct measurement. In CA space, the Plank distance ($1.6 \times 10^{-35}$ meters) scale roughly corresponds to the minimum event distance (or cell 'size'), and Plank time scale ($5.4 \times 10^{-44}$ sec) to the minimum time interval possible. We show that any CA model automatically leads to a <u>maximum speed</u> limit for the <u>transfer</u> of information from place to place in CA space, and hence leads to strict physical locality of all physical interactions. Furthermore, we show that the Lorentz transformation follows mathematically from CA theory, where photon propagation is postulated to be the simply shifting of the photon information pattern from cell to adjacent cell in every CA 'clock cycle' (which incidently is simplest possible CA 'motion'). Minkowski 4D flat space-time of special relativity can be seen as the direct consequence of the low level universal CA processes, as seen by inertial observers with ordinary measuring instruments, and who are not aware of, and cannot measure with the true absolute units of CA space and time.*

# TABLE OF CONTENTS

## 2. INTRODUCTION

*"Experiment has provided numerous facts justifying the following generalization: absolute motion of matter, or, to be more precise, the relative motion of weighable matter and ether, cannot be disclosed. All that can be done is to reveal the motion of weighable matter with respect to weighable matter…"*

*H. Poincare (1895)*

It has been over 90 years since Albert Einstein first conceived Special Relativity (SR). Since then, the basic postulates of relativity have been repeatedly verified experimentally (ref. 22) to a high accuracy. In fact, modern physics would be totally inconceivable without the pillar stone that special relativity provides. So why should anyone pay attention to yet another formulation of special relativity?

In spite of 90 years of success, the foundation of SR seems disturbing to some physicists. For example, Lorentz (one of the physicists who helped develop special relativity in the early 1900's) always remained skeptical about SR, and still believed in the existence of the ether till his death. Even today there are physicists who still question the foundations of special relativity (ref. 23). Some of the most disturbing results from SR include the abolishment of the ether and the relative nature of 4D space-time. In SR, it still is not understood why our universe has a limiting speed for all motion (and as such still remains a postulate). To some the propagation of light still seems to be cloaked in mystery. It also seems strange to some that the mass of an object (which depends on the sum of the masses of the elementary particles that make up the mass) is variable with respect to different inertial observers (ref. 23). Although all these results still hold true in our new derivation, the origins of these results are understood within a new SR framework.

Here we present a new derivation of Special Relativity that is based purely on the quantum nature of matter and light, and on the force particle exchange paradigm of quantum field theory. It is also based on the postulate that our universe is a vast Cellular Automata (CA) simulation. We believe that Cellular Automata (CA) (references 2,3 and 4) forms the basis of a unified model of nature. Cellular Automata theory will be discussed fully in the next section.

Cellular Automata is currently the *best* model we have for understanding the 'machinery' of our universe. CA theory suggests that **all** the laws of physics should be the result of interactions that are strictly local, which therefore forbids any action at a distance. If CA theory is correct, then the global laws of physics (like the Newton's law of inertia 'F=MA') should be the result of the local actions of matter particles, which exist directly as information patterns on the cells of the CA. In addition, CA theory suggests that space, time, matter, energy and motion are all the **same** thing, namely the result of information changing state according to some set of specific mathematical rules! We will see that Einstein's special relativity is already manifestly compatible with the cellular automata model of our universe. We have also developed a new theory of gravity called ElectroMagnetic Quantum Gravity (or EMQG, reference 1) which is a quantum theory of gravity, which is also manifestly compatible with the Cellular Automata model. This theory extends the concepts presented here to include acceleration and gravity.

If CA theory is true, this implies that our familiar notions of space, time, and matter are not the basic elements of physical reality. In CA theory, all physical phenomena turn out to be the end result of a vast amount of numerical information being processed on a 'universal computer', and that the information *inside* this computer is in fact our *whole* universe, including ourselves! All elements of reality are due to numerical information being processed on the 'cells' at incredibly 'high speeds' (with respect to us) on this universal computer. The computer hardware is forever inaccessible to us, because we ourselves are also information patterns residing on the cells. The laws of physics that govern the 'hardware' functioning of the universal computer hardware do not even have to be the same as our own physical laws. In fact, the computer 'hardware' that governs our reality can be considered by definition of the word 'universe' to be outside our own universe, because it is ***inaccessible.***

The computer model that best fits the workings of our universe is quite different from our ordinary personal desktop computers. In fact, our universe is implemented on the most massively parallel computer model currently known to computer science. This parallel computer model is called a 'CELLULAR AUTOMATA'. Not only is the CA a parallel computer, but it is also the fastest known parallel computer processor. A CA consists of a huge array of 'cells' (or memory locations) that are capable of storing numerical data, which change state on every clock period, everywhere, according to the rules of the cellular automata. Figure 1 shows a schematic of a 3D CA, which is the model proposed for our universe. This type of computer was discovered theoretically by Konrad Zuse and Stanislav Ulam in the late 1940's, and later put to use by John von Neumann to model the real world behavior of complex spatially extended structures. The best known example of a CA is the game of life originated by John Horton Conway. In fact, the CA is so powerful that it is capable of updating the entire memory (no matter how big) in a single clock pulse! Contrast this to the desktop computer, which takes millions of clock cycles to update the entire memory. CA's are inherently symmetrical, because one set of rules is programmed and repeated for each and every memory cell. We believe that this accounts for the high degree of spatial symmetry found in our universe. In other words, the laws of physics are the same no matter where you are, or how you are oriented in space. This observation is accountable by the perfect symmetry of the cell space.

We have postulated a simple model of photon propagation on the CA, which accounts for the known behavior of light. Light 'moves' (motion is really the shifting of patterns from place to place) in a kind of absolute, quantized 3D space, and separate 1D time (plank units) of the cellular automata. Photons simply shift from cell to adjacent cell, on each and every CA clock cycle. This motion is totally de-coupled from, and *unaffected by the source motion*. The light source only determines the energy, and therefore the wavelength of the light. Once the light leaves the source, the wavelength and velocity is an absolute constant, which is best specified in absolute CA units. We show from first principles (section 9) that the *measured* light velocity is constant in *all* space directions as measured by all inertial observers, which still remains only as a postulate of special relativity. We will show that the Lorentz transformation follows mathematically from this model. The Lorentz transformation is at the core of special relativity, and yields all the familiar results of special relativity such as: Lorentz time dilation, Lorentz length contraction, relativistic velocity addition formula, and so on.

On incredibly small distance and time scales called the Plank scale, (about $10^{-35}$ meters distance and $10^{-43}$ seconds, ref. 10) our 'view' of the universe is completely unlike what we know from our ordinary senses. Figure 1 shows what space might 'look' like at the lowest possible scales. Space, time, and matter no longer exist. Elementary particles of matter reveal themselves as oscillating numeric information patterns. Motion turns out to be an illusion, which results from the 'shifting' of particle-like information patterns from cell to cell. However, there is no real movement! Forces result from vector boson particle exchanges, which can be viewed as the exchange of oscillating information patterns, readily emitted and absorbed by matter particles (fermions). The absorption of a vector boson information pattern changes the internal oscillation of a particle, and causes an 'acceleration' to occur along a particular direction (usually towards or away from the source). The quantization of space reveals itself as cells, or storage locations for the numbers of the computer. What causes the numbers to change state as time progresses? It's the logical (and local) rules that are preprogrammed in all the storage cells of the computer. All the cells change state at the same time (not our time, but during a CA 'clock transition') and at regular CA 'clock' intervals (not to be confused with clocks in our universe). So, in CA theory at the lowest level, space isn't 'nothing', it is something; it's memory cells. Particle information patterns (numbers) residing in the cells are dynamic, and shifting (but ***not actually*** moving)! They simply change state as the computer simulation evolves. In order to understand CA theory and its relationship to modern physics we need to take a brief review of CA computer theory.

3.     CELLULAR AUTOMATA THEORY

*"Digital Mechanics is a discrete and deterministic modeling system which we propose to use (instead of differential equations) for modeling phenomena in physics.... We hypothesize that there will be found a single cellular automaton rule that models all of microscopic physics; and models it exactly. We call this field Digital Mechanics."*

*- Edward Fredkin*

In this section, we give a brief summary of Cellular Automata theory and it's connection with modern physics (ref. 2,3 and 4). We also include some **new** material we originally developed in regards to the CA space dimensionality. Edward Fredkin (ref. 2) has been first credited with introducing the idea that our universe is a huge cellular automata computer simulation. The cellular automata (CA) model is a special computer that consists of a large number of cells, which are storage locations for numbers. Each cell contains some initial number (say 0 or 1 for example), and the same set of rules are applied for *each* and every cell. The rules specify how these numbers are to be changed at the next computer 'clock' interval. Mathematically, a 'clock' is required in order to synchronize the next state of all the cells. The logical rules of a cell specifies the new state of that cell on the next 'clock' period, based on the cells current state, and on that of all the cell's immediate neighbors (each cell in CA has a fixed number of neighboring cells). The number of neighbors that influence a given cell is what we call the connectivity of the cellular automata. In other words, the number of neighbors that connect (or influence) a given cell is called the CA connectivity. The connectivity can be any positive integer number.

We define a 'geometric' cellular automata as a CA configuration where each cell has only the correct number of neighbors such that the CA connectivity allows a simple cubic geometric arrangement of cells (a cube in 3D space for example, see figure 1). This can be visualized as the stacking of cells into squares, cubes, hypercubes, etc. This structure is well suited for constructing 3D space, as we know it. The mathematical spatial dimensions required to contain the geometric CA is defined as the 'dimensionality' of the geometric cellular automata. For example, in a 2D geometric CA each cell has 8 surrounding neighbors, which can be thought of as forming a simple 2D space. One set of rules exists for every given cell, based on the input from its immediate 8 neighbors (and possibly on the state of the cell itself). The result of this 'computation' is then stored back in the cell on the next 'clock' period. In this way, all the cells are updated simultaneously in every cell, and the process repeats on each and every clock cycle. In a 1D geometric CA, each cell has 2 neighbors, a 2D geometric CA has 8 neighbors, a 3D geometric CA has 26 neighbors (figure 1), a 4D has 80 neighbors, and a 5D has 242, and so on. In general, if $C_D$ is the number of neighbors of an Nth dimensional geometric Cellular Automata, and $N_{D-1}$ is the number of neighbors of the next lower N-1th dimensional geometric Cellular Automata, then: $C_D = C_{D-1} + C_{D-1} + C_{D-1} + 2$, or $C_D = 3 C_{D-1} + 2$, which is the number of neighbors of an Nth dimensional geometric CA expressed in terms of the next lowest space.

The geometric CA model that is explored here for our universe is a simple geometric 3D CA, where each cell has 26 neighboring cells (figure 1). Your first impression might be that the correct geometric CA model would be a 4D geometric CA, so that it is directly compatible with relativistic space-time. There are several problems with this approach. First, 3D space and time have to be united in this CA model, which is not easy to do. More difficult still, is that 4D space-time is relative, and in some cases it is curved. This means that on the earth, an observer in a falling reference frame 'sees' flat 4D space-time, while a stationary observer exhibits curved 4D space-time. Furthermore, the 4D space-time curvature is directional near the earth. Curvature varies along the radius vectors of the earth, but does not vary parallel to the earth's surface (over small distances).

There are actually two space-time structures in our universe. First, there is the familiar relativistic 4D space-time that is measured with our instruments, and influenced by motion or gravity. Secondly, there is an absolute 'low level' 3D space, and separate 'time' that is not directly accessible to us, which exists at the lowest scale of distance and time, which we refer to as being on the Cellular Automata level. The 3D space comes in the form of cells, which are automatically quantized. Time is also automatically quantized, and is extremely simple. Our time is the unidirectional evolution of the numeric state of the CA based on the local rules on each 'clock cycle'. The numeric state of the CA changes state on each and every 'clock' pulse (an external synchronizer of the CA), which can be thought of as a primitive form of CA time ('clock cycles' should not to be confused with our clocks, or our measure of time). Thus, our low-level space and separate time is simply a 3D geometric CA. This model restores utter simplicity to the structure of our universe. Everything is deterministic, and the future evolution of the universe can in principle be determined, if the exact numeric state of the Cellular Automata is known at this point in time, along with the rules that govern each cell.

What is the scale of this 3D geometric CA computer with respect to our distance and time scales? Assuming that the quantization scale corresponds to the Plank Scale (ref. 10) the number of cells per cubic meter of space is astronomically large: roughly $10^{105}$ cells. (It turns out that the quantization scale is much finer than the Plank Scale of distance and time, according to EMQG). Remember that all the cells in the universe are all updated in one single 'clock' cycle! This is a massive computation indeed! The number of CA 'clock' pulses that occur in one of our seconds is a phenomenal $10^{43}$ clock cycles per second (based on the assumption that the Plank distance of $1.6 \times 10^{-35}$ meters is the rough quantization scale of space, and the Plank time is $5.4 \times 10^{-44}$ seconds). Because of the remotely small distance and time scales of quantization, we as observers of the universe are very far removed from the low-level workings of our CA computer. Why do humans exist at such a large scale as to be remotely removed from the CA cell distance scale? The simple answer to this is that life is necessarily complex! Even an atom is remarkably complex. A lot of storage locations (cells) are required to support the structure of an atom, especially in light of the complex QED processes going on. It is not possible to assemble anything as complex as life forms without using tremendous numbers of atoms and molecules.

Chaos and complexity theory teaches us that simple rules can lead to enormous levels of complexity. We can see this in a simple 2D geometric CA called the game of life. Being a 2D geometric CA, there are 8 neighbors for each cell, which forms a primitive geometric 2D space, which can be viewed on a computer screen. Here the rules are very simple. The rules for the game of life are summarized below:

**Rules for the famous 2D Geometric CA - Conway's Game of Life**:

(1) If a given cell is in the one state, on the next clock pulse it will stay one when it is surrounded by precisely two or three ones among it's eight neighbors on a square lattice. If the cell is surrounded by more than three neighbors, it will change to zero; if fewer than two neighbors have a one, it changes to zero.

(2) If a given cell is in the zero state, on the next clock pulse: it will change to a one when surrounded by precisely three ones, otherwise for any other combination of neighbor states the cell will remain a zero.

The rules are simple enough for a child to understand, yet the game of life leads to an endless number of different patterns, and to significant complexity. We see gliders, puffers, guns, 'oscillating' particles with different rates of translation and spontaneous particle emission from some oscillating patterns. We have even seen a pattern that resembles a particle exchange process.

How many different 2D geometric cellular automata's can be constructed from all possible rules? This number is unimaginably large. For simple binary cells, with 8 neighboring cells there are 8+1 cells that influence a given cell (previous state of a cell can influence it's next state), which leads to $2^{512}$ possible binary combinations or approximately $10^{154}$ different CA's, of which the game of life is but one. In general, for an Nth Dimensional Geometric CA with (m) neighbors, there are $2^k$ possible rules available for the Cellular Automata, where $k = 2^{(m+1)}$. Assuming our universe is a simple 3D geometric CA, then there are $2^{134,217,728}$ possible rules to choose from! You can give up trying to find the rules that govern our universal CA by simple trial and error.

In the early 1900's Max Plank (ref. 10) defined a set of fundamental scales based on his then newly discovered quantum of energy h (E = hv). Three fundamental constants G (Gravitational Constant), h (Plank's Quantum of Action), and c (the speed of light) were assembled in a set of equations that define natural physical units independent of any man made units. The Plank length is 1.6 x $10^{-35}$ meters and the Plank time is 5.4 x $10^{-44}$ seconds. The Plank length has often been suggested as the fundamental quantization scale of our universe. This suggests that the 'size' of a cell in the CA computer is the Plank length, and that the Plank time is the period of the cellular automata 'clock' (the smallest possible time period for a change in the state of the CA). Note: The cells actually have no real size, since they represent storage locations for numbers. Instead, this represents the smallest distance that you can increment as you move along a ruler. (The actual quantization scale of the CA is much finer than the Plank Scale by some currently unknown amount).

Even in conventional physics, there is growing evidence accumulating that suggests that the plank units of distance and time somehow represent the quantization scale of space-time itself. In CA theory, quantization is automatic! (According to EMQG, the quantization scale of the universal CA is much finer than the Plank Scale). There are also some other physical units that can be derived from the three fundamental constants listed above which includes: Plank Energy, Plank Temperature, Plank Mass, Plank Speed, and Plank Wavelength, which also represent fundamental universal limits to these parameters.

All physical things are the result of CA processes; including space, time, forces and matter (particles). Although, the exact rules of the CA that is our universe is unknown at this time, some very general physical conclusions can be drawn from the CA model. For example, matter is constructed from elementary particles, which move in space during some period of time. Particles interact via forces (exchange particles) which bind particles together to form atoms and molecules. Quantum field theory tells us that forces are also particles called vector bosons, which are readily exchanged between matter particles (fermions), and that these exchanges cause momentum changes and accelerations that we interpret as forces. Elementary particles and forces on the CA consist of oscillating information patterns, which are numbers changing state dynamically in the cells of the CA. These numerical information patterns roam around from cell to cell in given directions.

The shifting rates of a particle or information pattern, relative to some other particle, is interpreted by us as the state of relative motion of the particle. As we have seen, particles can interact by exchanging particles, which are also information patterns. Exchange particles are readily emitted at a given fixed rate by the source particle, and absorbed by target particle. When these force particles are absorbed, the internal state changes, which we interpret as a change in the particle momentum. The result of this process is that the shifting rates or motion of the matter particle changes by undergoing a positive or negative acceleration with respect to the source. This is what we observe as a fundamental unit of force. Of course, when we observe forces on the classical scale, the astronomical number of particle exchanges occurring per second blurs the 'digital' impact nature of the force exchange, and we perceive a smooth force reaction. All 'motion' is relative in CA theory, since all cells are identical and indistinguishable. In other words, we cannot know the specific cell locations that a particle occupies.

The closest CA elements that correspond to our space and time are the empty cells and the clock cycles that elapse. But this correspondence is not exact, as we shall find. The cells, which are storage locations for numbers, really form a low-level basis of the physical concept of space. Because the information patterns can roam freely in various directions that are determined by the dimensionality of the CA, we interpret this freedom of motion as space. Similarly, while matter patterns are in motion, a definite time period elapses. We can only sense the elapse of time when matter is in motion, by the changing state of the CA. The ultimate cause of change is the CA clock, and the common rules that govern the cells. But, it is important to realize that the internal clock required for the CA to function is not the same as our measure of time in our universe. Our time is based on physical phenomena only. This fact is the origin of much confusion on the nature of time in physics. This important concept is investigated in EMQG theory (ref. 1).

From these and other considerations, CA theory restores a great unity to all of physics. Where there used to be different phenomena described by different physical theories, now there is only one theory. Furthermore, CA theory is not only able to describe the way the universe works, but it also allows us to understand *how* it works in detail. Is there any evidence in the current laws of physics to support the idea that our universe really is a Cellular Automata computer simulation? The following sections will provide some rather speculative and sometimes circumstantial evidence to support this position.

## 4. THE QUANTUM VACUUM AND IT'S RELATIONSHIP TO CA THEORY

*Philosophers:    "Nature abhors a vacuum."*

One might think that the vacuum is completely devoid of everything. In fact, the vacuum is far from empty. In order to make a complete vacuum, one must remove all matter from an enclosure. However, this is still not good enough. One must also lower the temperature down to absolute zero in order to remove all thermal electromagnetic radiation. However, Nernst correctly deduced in 1916 that empty space is still not completely devoid of all radiation after this is done. He predicted that the vacuum is still permanently filled with an electromagnetic field propagating at the speed of light, called the zero-point fluctuations (sometimes called vacuum fluctuations). This was later confirmed by the full quantum field theory developed in the 1920's and 30's. Later, with the development of QED, it was realized that all quantum fields should contribute to the vacuum state, like virtual electrons and positron particles, for example.

According to modern quantum field theory, the perfect vacuum is teeming with activity, as all types of quantum virtual particles (and virtual bosons or force particles) from the various quantum fields, appear and disappear spontaneously. These particles are called 'virtual' particles because they result from quantum processes that have short lifetimes, and are undetectable.

One way to look at the existence of the quantum vacuum is to consider that quantum theory forbids the absence of motion, as well as the absence of propagating fields (exchange particles). In QED, the quantum vacuum consists of the virtual particle pair creation/annihilation processes (for example, electron-positron pairs), and the zero-point-fluctuation (ZPF) of the electromagnetic field (virtual photons) just discussed. The existence of virtual particles of the quantum vacuum is essential to understanding the famous Casimir effect (ref. 11), an effect predicted theoretically by the Dutch scientist Hendrik Casimir in 1948. The Casimir effect refers to the tiny attractive force that occurs between two neutral metal plates suspended in a vacuum. He predicted theoretically that the force 'F' per unit area 'A' for plate separation D is given by:

$$F/A = -\pi^2 h c / (240 D^4) \quad \text{Newton's per square meter} \quad (\text{Casimir Force 'F'}) \quad (4.1)$$

The origin of this minute force can be traced to the disruption of the normal quantum vacuum virtual photon distribution between two nearby metallic plates. Certain photon wavelengths (and therefore energies) in the low wavelength range are not allowed between the plates, because these waves do not 'fit'. This creates a negative pressure due to the unequal energy distribution of virtual photons inside the plates as compared to outside the plate region. The pressure imbalance can be visualized as causing the two plates to be drawn together by radiation pressure. Note that even in the vacuum state, virtual photons carry energy and momentum.

Recently, Lamoreaux made (ref. 12) accurate measurements for the first time on the theoretical Casimir force existing between two gold-coated quartz surfaces that were spaced 0.75 micrometers apart. Lamoreaux found a force value of about 1 billionth of a Newton, agreeing with the Casimir theory to within an accuracy of about 5%.

Quantum Inertia theory depends heavily on the existence of the virtual particles of the quantum vacuum, and so we present other evidence for the existence of virtual particles (briefly) below:

(1) The extreme precision in the theoretical calculations of the hyper-fine structure of the energy levels of the hydrogen atom, and the anomalous magnetic moment of the electron and muon are both based on the existence of virtual particles. These effects have been calculated in QED to a very high precision (approximately 10 decimal places), and these values have also been verified experimentally. This indeed is a great achievement for QED, which is essentially a perturbation theory of the electromagnetic quantum vacuum.

(2) Recently, vacuum polarization (the polarization of electron-positron pairs near a real electron particle) has been observed experimentally by a team of physicists led by David Koltick. Vacuum polarization causes a cloud of virtual particles to form around the electron in such a way as to produce charge screening. This is because virtual positrons migrate towards the real electron and virtual electrons migrate away. A team of physicists fired high-energy particles at electrons, and found that the effect of this cloud of virtual particles was reduced, the closer a particle penetrated towards the electron. They reported that the effect of the higher charge for the penetration of the electron cloud with energetic 58 giga-electron volt particles was equivalent to a fine structure constant of 1/129.6. This agreed well with their theoretical prediction of 128.5. This can be taken as verification of the vacuum polarization effect predicted by QED.

(3) The quantum vacuum explains why cooling alone will never freeze liquid helium. Unless pressure is applied, vacuum energy fluctuations prevent its atoms from getting close enough to trigger solidification.

(4) For fluorescent strip lamps, the random energy fluctuations of the vacuum cause the atoms of mercury, which are in their exited state, to spontaneously emit photons by eventually knocking them out of their unstable energy orbital. In this way, spontaneous emission in an atom can be viewed as being caused by the surrounding quantum vacuum.

(5) In electronics, there is a limit as to how much a radio signal can be amplified. Random noise signals are somehow added to the original signal. This is due to the presence of the virtual particles of the quantum vacuum as the photons propagate in space, thus adding a random noise pattern to the signal.

(6) Recent theoretical and experimental work in the field of Cavity Quantum Electrodynamics suggests that orbital electron transition time for excited atoms can be affected by the state of the virtual particles of the quantum vacuum surrounding the excited atom in a cavity.

What relationships exist between CA theory and the quantum vacuum? Recall that the quantum vacuum implies that **all** of empty space is filled with virtual particle processes. In simple 2D geometric CA's (such as the Conway's Game of Life), most random initial states or 'seed' patterns on the cells (and often from small localized initial patterns with all the remaining cells in the zero state) often evolve into a complex soup of activity, everywhere. This activity is very reminiscent of our quantum vacuum. In the game of life you can even see events that even look suspiciously like random 'particle' collisions, particle annihilation, and particle creation after a sufficiently long period of simulation time. Of course this is not hard evidence for CA theory, but it is never the less highly suggestive.

## 5.     THE BIG BANG (START OF SIMULATION AT T=0) AND CA THEORY

*"I want to know how God created this world (Universe)"*              *- A. Einstein*

*"Nothing can be created out of nothing"*                             *- Lucretius*

If the universe is a CA computer simulation, then there must have been a point where the simulation was first started. This occurred ≈15 billion years ago (our time), according to the standard big bang theory. It is important to realize that the creation of the numeric state of our universe (if it were to be done *now,* as a single act of creation without evolution) would be **very** difficult to accomplish now. All the galaxies, stars, and planets, and life forms must be specified for all the states of the cellular automata cells, which for our universe is something on the order of something like $10^{100}$ cells per cubic meter of space!

Our universe contains on the order of a few billion galaxies, and many galaxies have on the order of 100 billion stars in it. Currently, there is also evidence for the possibility that a certain percentage of these stars have one or more planets circling around them. Each star and planet has it's own unique orbit, chemical composition, temperature, rotation rate, size, atmosphere, landscape and possibly even life forms. In the process of creating our universe, it is far more economical to start with just the "right" rules of the cellular automata so that stars and planets are the natural byproducts of the evolution of the CA (and possibly life as well). In other words, let the natural evolution of the CA run its course. It is also more "interesting" to start this process, and than "see" what comes out of it after a lot of computer processing. In fact, that is what the purpose of our universal CA computer is, it is to compute our universe! CA theory absolutely requires that our universe be an evolutionary process, with a simple beginning.

## 6. WHY OUR UNIVERSE IS MATHEMATICAL IN NATURE

*"Why is it possible that mathematics, a product of human thought that is independent of experience, fits so excellently the objects of physical reality."*

*- Albert Einstein*

It is clear that all the known laws of physics are mathematical in nature. Many physicists like Einstein, for example, have commented on this mysterious fact. No good explanation has been given as to why this should be so. This fact is made even more mysterious when one considers that mathematics is strictly an invention or byproduct of intellectual activity. In a sense, mathematics is like art and music. For example, the mathematical concepts of infinity, the imaginary numbers, and the Mandelbrot set in the complex plane are all mathematical objects that are invented by mathematicians. In mathematics, you start with virtually any set of self-consistent axioms, and formulate new mathematics as you please. Mathematics is strictly a *creative* process. Yet, our universe definitely operates in a mathematical way. Every successful physical theory has been formulated in the language of mathematics, and a good theory can even predict new phenomena that was not expected from the original premises.

The cellular automata model provides a clear explanation as to why the universe is mathematical. Quite simply, everything in our universe is numerical information, which is governed by mathematical rules that specify how the numbers change as the computation progresses. In short, "***the universe is numbers***"**,** as was once proclaimed by the great Greek philosopher and mathematician Pythagoras. The design of the cellular automata must have required intelligence, which was applied to the cellular automata in the form of the mathematical rules for the cells. CA theory claims that all the laws of physics that we know today are mathematical descriptions of the underling, discrete mathematical nature of the numeric patterns that are present in our universal cellular automata. Fredkin (ref. 2) once proposed that the universe should be modeled with a single set of cellular automaton rules, which will model all of microscopic physics exactly. He called this CA 'Digital Mechanics'. The laws of physics in this form are discrete and deterministic, and would replace the existing differential equations (based on the space-time continuum) for modeling all phenomena in physics.

Therefore, the current thrust to discover the theory of everything (or simply The Theory as it is now known) should not be looking for a set of partial differential equations incorporating relativity and quantum field theory. Instead, we should be looking for the correct structure of the universal CA and the set of logical rules that govern it's operation.

## 7. QUANTUM PARTICLES IN SAME STATE ARE INDISTINGUISHABLE

*" Common sense is the layer of prejudice laid down in the mind prior to the age of 18"*

*A. Einstein*

A particle physicist once remarked that elementary particles behave more like mathematical objects than like familiar point-like objects. Particles are able to transform from one species type to another. Particles seem to be spread-out in some sort of oscillatory wave, and at other times they seem like point-like objects. Particles can be readily annihilated and created. None of these processes seem familiar from our everyday experience. One of the most unfamiliar of all particle attributes is indistinguishability.

Quantum Mechanics teaches us that electrons in the same quantum state (or having the same quantum numbers) are absolutely identical, and indistinguishable from each other. You cannot mark one electron so that it is different than another. An electron is currently described by quantum mechanics as a particle with quantum numbers like: mass, charge, spin, position, and momentum, which are represented as numbers in the wave function of the electron. It is these properties alone tell you all there is to know about the electron. The electron has no size or shape. Quantum Mechanics has definitely ruled out any classical or 'mechanical' models to help us 'visualize' what an elementary particle really is.

Equality is strictly a mathematical concept. In mathematics, the equality 1+1=2 is exact. In classical physics, no two marbles can be constructed to be exactly the same. When it comes to elementary particles, however, two quantum particles can be *exactly* the same. According to quantum mechanics, two electrons in the same state of motion (and spin) are absolutely identical and indistinguishable. The cellular automata model explains this remarkable fact simply by stating that the two electrons in the same state have *exactly* the same numeric information pattern, and thus described by the same quantum wave function. Therefore, they are mathematically identical. In constructing a universe, it is very desirable to have building blocks that are identical, and exactly repeatable, so that large complex structures can be easily formed.

## 8. EINSTEIN'S RELATIVITY IN THE CONTEXT OF CA

It is important to note that ***special relativity is already manifestly compatible with our CA model***! In other words, we have found that the basic postulates and their mathematical consequences agree well with CA theory. However, the interpretation of forces, acceleration, and the famous mass-velocity variation requires updating, as we shall soon see.

We have proposed that our universe is a vast Cellular Automaton. For this to be true, all physical phenomena must come from the strictly local interactions inside the CA. The very nature of the cellular automata model is totally incapable of any instantaneous action at a distance, since information can only be sent from cell to adjacent cell in any direction, only at each and every 'clock' pulse. This means that there can be no action at a distance in *all* the laws of physics. Einstein abolished action at a distance in special relativity with his famous velocity of light postulate. He also removed gravitational action at a distance in general relativity by replacing Newton's instantaneous gravitational force law with his space-time curvature concept. In the following sections, we will see how and why special relativity is manifestly compatible with our CA model.

Special relativity is one of the most successful theories of all physics, and along with quantum theory forms one of the two great pillars of modern physics. However, it has failed to account for *why* the universe has a maximum speed, which has still remained as one of the two postulates of special relativity. CA theory provides a simple explanation for this. In fact, the CA model *demands* that the universe have a maximum speed limit! In addition to this, the second postulate regarding the relativity of inertial frames (constant velocity motion) can also be seen as a simple consequence of the basic structure of the CA. The focus of the rest of the paper is on reformulating special relativity with these new principles in mind.

General relativity as it is currently formulated, is *not* compatible with CA theory. First, general relativity is formulated with the classical continuum concept for matter-energy, and is also formulated with a special space-time continuum defined by a metric tensor. Both of these fields are not generally compatible with the CA model, or with quantum theory in general. Secondly, there is no known local action that couples a large mass to the surrounding space-time curvature. What is it about a large mass that causes space-time curvature around it? In general relativity, there exists a global tensor field called the 4D space-time metric, which merely describes the amount of the curved 4D space-time. However, the relative nature of space-time (observers in free fall near the earth live in flat space-time), makes it very difficult to conceive how relativistic 4D space-time can actually work on a CA. How does the principle of equivalence work on a CA? Why does the inertial mass equal the gravitational mass, especially since they are defined differently?

General relativity has also failed to make any progress towards the understanding of inertia. Inertia is introduced in general relativity exactly as it was conceived by Newton in his famous inertia law: F=MA. Associated with Newton's formulation of inertia are the problems introduced by Mach's principle, which is a loose collection of ideas and paradoxes that have to do with accelerated or rotating motion. Mach argued that motion would appear to be devoid of any meaning in the absence of some surrounding matter, and that the local property of inertia must somehow be a function of the cosmic distribution of all the matter in the universe. Mach's principle has remained as an *untestable* philosophical argument, even within the scope of general relativity.

We found that general relativity must be revised in order to be compatible with CA theory. These modifications of general relativity came about by our new understanding of inertia and the principle of equivalence. Inertia results from low-level quantum processes that is compatible with CA theory, which also resolves Mach's paradox. This new theory of inertia is called Quantum Inertia (QI). QI also helps in explaining the origin of the Einstein Principle of Equivalence (the weak version). Equivalence is ***not*** a fundamental principle of nature, but turns out to be due to similar quantum processes occurring in accelerated frames and gravitational fields. The reformulation of general relativity to incorporate Quantum Inertia, the quantum origins of the Principle of Equivalence, and CA theory is called 'ElectroMagnetic Quantum Gravity' or EMQG (reference 1). It is a quantum theory of gravity because matter is treated as quantum particles and 4D space-time is quantized (results from pure quantum particle processes). Furthermore, the quantum action that occurs between a large mass and the surrounding 4D space-time is clearly understood in EMQG. However, we will focus only on special relativity in this work (refer to reference 1 for the details of EMQG theory).

## 9. SPECIAL RELATIVITY AND CELLULAR AUTOMATA

*"... space by itself and time by itself, are doomed to fade away into mere shadows ...."*

- H. Minkowski

**THE BASIC POSTULATES OF SPECIAL RELATIVITY**

Special Relativity theory is founded on two basic postulates:

**(1) The velocity of light in a vacuum is constant and is equal for all observers in inertial frames (inertial frame is one in which Newton's law of inertia is obeyed).**

**(2) The laws of physics are equally valid in all inertial reference frames.**

These postulates are used by Einstein to derive the famous Lorentz transformations, a set of equations that relate space and time measurements between different inertial frames. The second postulate implies that there are no absolute reference frames in the universe that can be used to gauge constant velocity motion. All inertial frames are equally valid in describing velocities. In a general sense, all the laws of physics are also equally valid in all inertial frames. Some of the important consequences of special relativity and the Lorentz transformations are:

(1) The universe is four dimensional, where 3D space and time now have to be united.
(2) There is a maximum speed to which matter can obtain.
(3) Mass and energy are interchangeable.
(4) Momentum (and mass) is relative. Mass varies with the relative velocity between two inertial frames.
(5) Spatially separated events that are simultaneous in one inertial frame are not generally simultaneous in another inertial frame.

The special theory of relativity also implies that the speed of light is the limiting speed for any from of motion in the universe. Furthermore, light speed appears constant no matter what inertial frame an observer chooses. However, nowhere in special relativity theory, (or any other theory we are aware of) is there an explanation as to why this might be so. It is simply a postulate, based on physical observations such as the Michelson-Morley experiment. The second postulate also implies that there are no experiments that can be performed that will reveal which observer is in a state of 'absolute rest'.

The second postulate of special relativity states that the laws of physics are equally valid in all inertial reference frames. Stated in a weaker form, there are no preferred reference frames to judge absolute constant velocity motion (or inertial frames). This latter form is easily explained in CA theory, by remembering that all cells and their corresponding rules in the cellular automata are absolutely identical everywhere. Motion itself is an illusion, and really represents information transfers from cell to cell. To assign meaning to motion in a CA, one must relate information pattern flows from one numeric pattern group with respect to another group (the actual cell locations are inaccessible to experiment). Therefore, motion requires reference frames. Unless you have access to the absolute location of the cells, all motion remains relative in CA theory. In other words, there is *no* reference frame accessible by *experiment* that can be considered as the absolute reference frame for constant velocity motion. (Later, we will see that virtual particles of the quantum vacuum still do not allow us to reveal our (constant velocity) motion between two inertial frames. However, this is not the case for an accelerated observer or for observers in gravitational fields).

Since the contents of the cells and their locations are not physically observable to us, they cannot be used to help us setup a universal absolute reference frame for motion. However, there does exist a universal reference frame in the CA, and this frame is completely hidden from experimentation, which we call the 'CA absolute reference frame'. We associate with this frame an *absolute space* and *absolute time*. Everyday objects like this desk, which is a very large collection of elementary particles, occupies a specific volume of cells in CA space. These cell patterns are (most likely) shifting through our cell space at some specific rate. Therefore, there does exist a kind of Newtonian absolute space and absolute time scale, but these are hidden from the viewpoint of an observer living in the CA. The idea of absolute CA space and CA time becomes very important in considerations of inertial and gravitation frames. Even more important to observers, is the state of the virtual particles of the quantum vacuum. These virtual particles can act to produce forces for observers in a state of acceleration. Mach's principle and Einstein's weak principle of equivalence depend on the existence of virtual particles.

## 9.1   THE NATURE OF LIGHT ON THE CELLULAR AUTOMATA

Imagine turning on a light switch. The photons are emitted from the tungsten filament and propagate to your eyes. Do the photons from the filament accelerate from some resting state, and rapidly stabilize to the fixed light velocity (say within a fraction of the size of the atom), or does the photon immediately move at light velocity from creation? Theoretically, this is a question that can be only answered by QED (Quantum Electrodynamics). In QED, a particular photon is emitted from an electron in the filament, which is excited to a higher energy level by the flow of current. The actual photon emission process can be traced to a fundamental QED vertex (one of the 8 fundamental processes defined by Feynman). When the photon is emitted from an electron, it causes a change in momentum (or release of energy) from the emitting electron. QED is completely based on special relativity, and therefore the photon velocity is always constant after emission. Since QED has been tested experimentally to remarkable accuracy, we can assume that there is some truth in this.

Therefore, light seems to be fundamentally different than matter. All observations of light velocity indicate that the velocity is always a constant, and therefore photons are created at light velocity. In the motion of matter, elementary matter particles can achieve any velocity (below light velocity). If matter moves from a standing state, it must always accelerate to achieve a given velocity. Why does photon motion differ in such a fundamental way from the motion of matter particles? The CA model provides a simple explanation for this. We will find that the photon motion represents the simplest possible 'motion' on the CA. We will see that the photon information pattern simply moves from cell to adjacent cell at every CA 'clock cycle', without regard to the motion of the source of the photon. Furthermore, the photon moves at only *one* fixed speed! No other velocities are possible for the photon. This agrees well with the apparent nature of light in special relativity, and in QED.

In any cellular automata, the clock rate specifies the time interval in which all the cells are updated, and acts as the synchronizing agent for the cells. Matter is known to consist of atoms and molecules, which themselves consist of elementary particles bound together by forces. An elementary particle in motion is represented in CA theory by a shifting numeric information pattern that is free to 'roam' from cell to cell. Recall that space consists of cells or storage locations for numbers in the cellular automata, and particles (number patterns) freely 'move' in this cell space. From these simple ideas, it can be seen that there must be a maximum rate that number patterns are able to achieve. This is due to the following two reasons. First, there is fixed, constant rate in which cells can change state. Secondly in CA theory, information can only be transferred sequentially, from one cell to adjacent cell, and only one cell at a time per clock cycle. This is simply a limitation of the structure of the cellular automata computer model. The CA structure provides the most massively parallel computer model known. It is the CA's high degree of parallelism that is responsible for this limitation, because a particular cell state can only be affected by its immediate neighbors. Information can only evolve after each 'clock' period, and information can only shift from cell to adjacent cell. This results in a *definite* **maximum speed limit** for information transfers on the CA.

*(NOTE: In EMQG this maximum speed actually represents the raw or 'low-level' light velocity, defined as the velocity of light in between encounters with charged virtual particles. The scattering of photons with the virtual particles of the quantum vacuum complicates everything, and reduces the speed of the photons to the familiar observable light velocity through a process known as photon scattering, reference 1).*

This maximum speed limit can be calculated if the precise quantization scale of space and time on the cellular automata level is known. Let us assume for now that the quantization of space and time corresponds *exactly* to the plank distance and the time scales. This means that the shifting of one cell represents a change of one fundamental plank distance $L_P$: $1.6 \times 10^{-35}$ meters, and that the time required for the shift of one cell is one fundamental plank time $T_P$: of $5.4 \times 10^{-44}$ seconds. Let us further assume that a photon represents the fastest of all the information patterns that shifts around in the CA. In fact, we propose that the photon information pattern is *only* capable of shifting one cell per clock period, and not at any other rate, and therefore exits at one speed with respect to the cells (figure 2). The value for the speed of light can then be derived simply as the ratio of (our) distance over (our) time for the information pattern transfer rate. The maximum information transfer velocity is thus:

$$V_P = L_P / T_P = 3 \times 10^8 \text{ meter/sec} = c \qquad (9.1)$$

Therefore, $V_P = c$, the speed of light. The velocity of light can also be expressed as one plank velocity, which is defined in units of plank length divided by plank time. (There are plank units for mass, temperature, energy, etc as detailed in ref. 10).

Thus, the fastest rate that the photon can move (shift) is an increment of one cell's distance, for every clock cycle. If two or more clock cycles are required to shift information over one cell, then the velocity of the particle is lower than the speed of light.

To summarize, in cellular automata theory the maximum speed simply represents the *fastest* speed in which the cellular automata can transfer information from place to place. Matter is information in the cellular automata, which occupies the cells. The cells themselves provide a means where information can be stored or transferred, and this concept corresponds to what we call the 'low level' discrete space. 'Low level' time corresponds to the time evolution of the state of the cellular automata, which is governed by the 'clock period'. To put it another way, the rate of transfer of information in any cellular automata is limited, and infinite speeds are simply not possible. Of course, this rules out action at a distance, which is why CA theory is manifestly compatible with special relativity.

In passing, it is interesting to note that in the famous 2D Geometric CA, called Conway's game of life, there exists a stable, coherent 'L' shaped pattern commonly known as a 'glider' pattern. This pattern is always contained in a 3 x 3 cell array, and the glider completes a kind of an internal 'oscillation' in four clock cycles. Thus, in four clock cycles it returns to it's initial 'L' shaped starting pattern. This glider travels in 2D cell space, at *one fixed speed*! It is also the fastest moving pattern known in Conway's game of life. The glider particle in some sense resembles the photon particle in our universe! It has an internal oscillation, and it only moves at one fixed velocity. However, the similarity ends here, because in the game of life, the glider only moves in four fixed directions.

9.2    THE LORENTZ TRANSFORMATION AND CA THEORY

Special Relativity predicts through the work of Minkowski, that space and time must be united into a special four-dimensional space-time structure. This unification can be thought of as being the only way to restore the mathematical concept of space in special relativity. In this way, observers in any state of uniform motion can agree on the measure of a mathematical 'distance' between two events. According to special relativity, if a distance "d" is measured in frame A, then this distance is generally different than that measured in another frame B which is in the state of constant velocity motion 'v' with respect to frame A. As a result, the ordinary 3D formula for the distance between two points $(x_1,y_1,z_1)$ and $(x_2,y_2,z_2)$ at time t, which is given by $d^2 = (x_2-x_1)^2 + (y_2-y_1)^2 + (z_2-z_1)^2$ will generally not be agreed on by all observers. This is because of the special relativistic distance contraction and time dilation that occurs in co-moving frames. But Minkowski showed that if a four dimensional coordinate system is properly chosen with the following line element: $ds^2 = dx^2 + dy^2 + dz^2 - c^2 dt^2$, (or which is represented by a 4D coordinates given by (x,y,z,ict) where $i=\sqrt{-1}$) then distance and time measurements become invariant under any inertial frame of reference. In other words, all inertial observers agree on measurements of space and time in this coordinate system regardless of the relative velocity. This is what we refer to as the high level 4D space-time continuum of special relativity. This is the space-time that is accessible to measurement, and generally depends on the state of relative motion of the observer. Most importantly, this space-time system is *relative*, and depends on the state of motion of an observer.

It can be shown that the constancy of light velocity, and the principle of relativity (Einstein's first and second postulate) leads directly to the famous Lorentz Transformations, a set of equations that allows us to relate space and time measurements between two different inertial reference frames. The Lorentz transformations between an observer 'B' in (x*,y*,z*,t*) moving at velocity 'v' with respect to the reference observer 'A' in (x,y,z,t) are given by:

$$x^* = (x - vt) * (1 - v^2/c^2)^{-1/2} \quad (9.2)$$
$$y^* = y$$
$$z^* = z$$
$$t^* = (t - (v/c^2) x) * (1 - v^2/c^2)^{-1/2}$$

These equations are derived by examining the motion of light that is spreading in a spherical wave front, starting at time t=0 (in the frame of observer 'A'), from the perspective of both inertial observers. Based on the constancy of light velocity in all inertial frames, the equations for the spherical wave front at time t for both observers (recall that the equation of a light sphere is: $x^2 + y^2 + z^2 = r^2$, where r=ct and r=ct* and the velocity of light is the same for both observers) is given by:

$$x^2 + y^2 + z^2 = c^2 t^2 \quad \text{... light sphere seen by observer 'A' after time t} \quad (9.3)$$
$$x^{*2} + y^{*2} + z^{*2} = c^2 t^{*2} \quad \text{... light sphere seen by observer 'B' after time t*} \quad (9.4)$$

It is a matter of pure algebra (which we will not repeat here, see ref. 20 for the detailed calculations) to derive the Lorentz transformation from these equations. The Lorentz transformation is at the heart of special relativity, from which we can derive the relativistic velocity addition formula, Lorentz length contraction, Lorentz time dilation, and many other results. Notice that the light velocity 'c' in the above equations is the measured light velocity, measured with identically constructed rulers and clocks by both observers.

## 9.3   MEASUREMENTS IN ABSOLUTE CA SPACE AND TIME UNITS

How can we translate the behavior of photons existing in CA absolute space and separate absolute time, into a statement concerning the *measurable* light velocity? In other words, into a statement based on an inertial observer's actual measurement data for his light velocity measurement. Furthermore, how do we compare these readings with respect to other inertial observers, who also use *actual measuring instruments*? A definition of a space and time measurement must be defined, along with a method of comparing these measurements among different observers in different inertial reference frames. This definition is required, because light velocity is defined as the measured distance that light moves, divided by the measured time that is required to cover this distance.

First we must define an inertial reference frame in far space, away from gravitational fields. Imagine a three dimensional grid of identically constructed clocks (figure 4), placed at regular intervals measured with a ruler, in the three dimensional space (ref. 19). Local observers are stationed at each of the clocks. Thus, the definition of an inertial frame is a whole set of observers uniformly distributed in space as we have described. All observers in a given reference frame agree on the position and the time of some event. Only one observer would actually be close enough to record the event (an event is defined as something that occurs at a single point in space, at a single instant in time). The data collected by all the observers are communicated to the others at a later time (by any means). Notice how light naturally enters in the definition of an inertial reference frame. Light is required by observers to literally 'see' the clock readings.

Now we are in a position to evaluate the Lorentz transformations from our low-level CA definition of space, time, and constancy of light velocity. Light is an absolute constant in absolute CA space and time units. No matter what the state of motion of the source, whether it is an inertial source of even if it is accelerated, the light moves as an absolute constant that is unaffected by the source. We must now translate this statement about *measured* light velocity, into the actual reality of the CA with imaginary observers with highly specialized measuring instruments capable of measuring plank distance and time units (which is not possible in our reality). Let us introduce an absolute, discrete (3D space) integer array: $[x(k),y(k),z(k)]$, where information changes state at every $t(k)$. These units represent our absolute space and time measurements (but in practice, we cannot actually make these measurements). The origin is an arbitrarily chosen cell (which can be looked at as being at absolute rest on the CA). A shift of data from one cell in any space direction to next, for example from $x(5)$ to $x(6)$, represents one plank distance unit (pdu), and if this take one clock unit, it happens in one plank time unit (ptu). The velocity of light represents one plank velocity unit (pvu) in our absolute units (figure 2). We intend to show that when two different inertial observers measure light velocity using *absolute* space and time units, both observers *measure* light velocity as being one plank velocity. However, space and time *measurements* between our two inertial observers, do *not* compare in our absolute units. We will show that this is the same situation we find in special relativity, for two observers with *real* measuring instruments in space-time.

Imagine two inertial observers with a relative velocity '$v_r$' in the CA absolute units (figure 3). Both observers are in a state of constant velocity motion with respect to our absolute cell coordinate system. Observer 'A' contains a green light source and moves with absolute velocity $v_a$ with respect to the cell rest frame. Observer 'B' moves with absolute velocity $v_b$, and is moving away from our observer 'A' (so that $v_b > v_a$), and $v_r = v_b - v_a$. Both observers carry measuring instruments capable of measuring space and time in absolute units. Of course, this is not actually possible with real observers.

Observer 'A' measures the velocity of light of his green light source, with his measuring instruments. He uses a ruler of length 'd' in absolute units, and measures the number of CA clock cycles it takes for the wave front of the green light to move the length of the ruler. Because observer 'A' is moving with velocity $v_a$, with respect to the absolute frame, his measurement of length and time are distorted (figure 3). Recall that light simply shifts from cell to cell, in every clock unit immediately after leaving the source. His measurement of length using the wave front moving across his ruler, appears longer, because of his motion $v_a$. Thus, observer 'A' distance measurement appears longer by: '$d + v_a d$' pdu, where $v_a$ is less than one. (for example, if $v_a$ = ½ pvu, and d=1,000,000 pdu, then the distance measured is 1,000,000 + ½ 1,000,000 pdu). In comparison, an observer at absolute rest would measure a distance of 'd' pdu. Similarly, the clock measures a longer time, because it takes longer for the wave front to reach the end of the receding ruler. Therefore, the time required to transverse the ruler is: '$d + v_a d$' ptu (in our example, the time taken for light to traverse the ruler is 1,000,000 + ½ 1,000,000 ptu). Thus, the *measured* light velocity in absolute CA units is: $(d + v_a d) / (d + v_a d) = 1$ pvu, the velocity of light. Similarly, for observer 'B' moving at velocity $v_b$, the measured velocity of the green light he receives in his reference frame is: $(d + v_b d) / (d + v_b d) = 1$ pvu, again equal to light velocity in absolute units. Thus, both observers conclude that light is a universal constant, equal to one pvu, no matter what the state of motion of the light source in an inertial frame! This is similar to the same situation in ordinary space-time.

What happens if observer 'A' sends his measurements (figure 3) to observer 'B' (by any means, carrier pigeon for example)? First, will observer 'B' conclude that the color of light received from 'A' is green? Secondly, will the measured distances and times be equal? It is obvious from the above analysis, that the measurements are not equal, unless $v_a = v_b$! Furthermore, observer 'B' concludes that the received light is shifted towards the red. Why? Observer 'B' examines the light received from 'A'. A 'wave marker' passes by him, and he then finds that the next 'wave marker' appears to take a longer time to arrive, compared to when both observers are both at absolute rest. Thus, the light appears to have a longer wavelength that is shifted towards the red, when compared to observer 'A'. The actual spacing between 'wave markers' is constant, and was determined by the energy of observer 'A's light emitting equipment. Note that observer 'A's measurement of his light wavelength at velocity $v_a$ is actually different from the wavelength measurement when he is at absolute rest, when measured in absolute units!

## 9.4 RELATIVISTIC MASS VARIATION FROM PARTICLE EXCHANGES

Let us now examine the results of the same experiment with measurements made in *ordinary* space-time, with ordinary measuring equipment like clocks and rulers. A common reference is required to make comparison measurements of length and time, since the absolute coordinate system is *not* available. Based on our definition of reference frames (as a grid of observers), light becomes the natural choice for comparative space-time measurements. Observer 'A' decides to define length in terms of the green light from his light source, where one basic length unit (bdu) ≡ 1000 wavelengths of green light, from which he has constructed a standard ruler of this length. Similarly, observer 'A' chooses to define the time of one basic time unit (btu) as the elapsed time required to receive 1000 cycles (or 1000 audible clicks from each wave crest, for example) of the green reference light, from which he constructs a calibrated standard clock. Observer 'B' has the identically constructed ruler and clock. Now, as before, observer 'B' has a relative velocity of $v_r$, with respect to 'A'. What happens when observer 'B' makes measurements on the incoming green light, sent by observer 'A'?

Now we do *not* have the luxury of absolute units to arbitrate between the two observers. Furthermore, no observer can be regarded as being at absolute rest! Both observers have an equal right to formulate the laws of physics of motion in his own frame. Observer 'A' measures the light velocity as follows: The green light travels distance '$D_a$' in time $T_a$, and therefore the measured light velocity is: $c = D_a/T_a$. Observer 'B' uses his identically constructed standard clock and standard ruler to measure the incoming green light. Does his measuring instruments measure the velocity of light the same as observer 'A'. The answer is yes. Recall that observer 'A''s velocity does not affect the light velocity at all. It is an absolute constant, and cannot be affected by the source motion. Recall that observer 'A' specifies the wavelength of light, through his source apparatus. Once set, the wavelength of light propagates as a constant, not affected by the source (as described in our CA model of light above). Therefore, observer 'B' measures the light velocity as follows:

$$(D_a + kD_a) / (T_a + kT_a) = [(1 + k)/(1+k)] (D_a/T_a) = D_a/T_a = c \text{ as for observer 'A'.} \qquad (9.5)$$

The motion of observer 'B' ruler adds a length of $kD_a$ cycles of light to his measurement distance, and adds the same $kT_a$ time delay, leaving the measured light velocity the same as 'A'. Observers 'A' and 'B' decide to compare their space and time measurements, with their identically constructed ruler and clock. Do these measurements agree? It is very clear that they do not!

Observer 'B' performs similar measurements, with identically constructed equipment on the incoming green light. Observer 'B' notices that the wavelength of the green light is shifted to the red, as we just discussed. Thus, his standard ruler of a length of one bdu contains *less* than 1000 wavelengths of the incoming light, because each wavelength is longer than 'A's (recall that 1 bdu is the length containing 1000 wavelengths of light). Similarly, he notices that when he listens for 1000 audible clicks (which should correspond to one btu of time), more than one btu of time elapses on his identically constructed clock (because each click takes a longer time to arrive). When the results of these measurements are compared by any means (by carrier pigeon, for example) observer 'B' concludes that his time has been dilated, and his distances have contracted compared to observer 'A's measurement. Incidentally, if observer 'B' has the green light source and shines it towards 'A', observer 'A' would conclude the same thing. How do we mathematically compute the values of these space-time comparisons? One may be tempted to apply a one dimensional Doppler-type analysis to deduce the quantity of space-time distortion. This, however, would *not* yield the correct answer. The above analysis is applicable for all the 3 dimensions that light can travel in space. Therefore, one must correct for light moving in all directions. This is precisely how Einstein derived the Lorentz transformations! In other words, the velocity of light measured in all directions of an expanding spherical wave front is what we take to be a constant. Thus, by showing that the velocity of light propagates as a constant in all directions in CA absolute space and time, we find that all inertial observers measure light velocity as a constant. However, they do not agree on the actual values of the space and time measurements. In this way, the principle of relativity leads us directly to the Lorentz transformation.

In summary, by postulating that on the lowest level of the CA, photons are information patterns 'moving' by a simple shifting from cell to adjacent cell at every clock 'cycle' in any given direction. As a consequence of this we found that:

(1) Light propagates in a kind of absolute, quantized 3D space, and separate 1D time (plank units) of the cellular automata, whose velocity is totally *unaffected by the source motion*. The light source determines the energy, and therefore the wavelength of the light. Once the light leaves the source, the wavelength and velocity is an absolute constant, specified in absolute CA units.
(2) In absolute CA space and CA time units, observers have an absolute velocity. The actual cell addresses of the information on the CA form the absolute 'rest' frame (which is not directly accessible by experiment). Hypothetical measurements in these absolute units yield light velocity and wavelength to be a constant, no matter what the state of motion of the source.
(3) When two (or more) inertial observers, with real measuring instruments are employed, and the measurements are made in the familiar 4D space-time defined by relativity theory, we have shown that all observers *measure* the velocity of light as a constant. However, when two (or more) inertial observers compare their space and time measurement (which is required to measure velocity, the measurements can be communicated by any means), they find that the measurements do not agree.

(4) We showed that the *measured* light velocity is constant in *all* space directions, which still remains only a postulate of special relativity. The Lorentz transformation directly follows from this through simple algebra. The Lorentz transformations form the core of special relativity, and yields the familiar results of relativity such as: time dilation, Lorentz contraction, velocity addition, and so on.

In regards to inertial frames, one might be tempted to consider that the virtual particles of the quantum vacuum might act as some sort of an abstract universal reference frame. One might think that the virtual particles in the neighborhood of a point might be used to gauge your constant velocity motion, a frame that would have been unknown to Einstein when he formulated special relativity. However, the virtual particles have completely random velocities, move in completely random directions, and most importantly are short lived and unobservable. Furthermore, one cannot 'tag' the virtual particles with labels, and follow the progress of all the virtual particles in order to judge your own motion with respect to the average motion of the virtual particles! Therefore, it is impossible to tell your state of constant velocity motion with respect to the vacuum, unless a force or some other vacuum phenomena makes it's presence felt. It is a well-known experimental fact that virtual particles introduce no new forces for inertial observers. However, this is definitely not the case for an accelerated frame, where we are concerned with the state of the acceleration vectors of the virtual particles with respect to a Newtonian accelerated mass (F=MA). Here forces are present, which originate from the electromagnetic interaction of the quantum vacuum with the matter.

Acceleration is a special motion, because an accelerated observer can detect his state of acceleration (inside a closed box, for example) by simply measuring the force exerted on him with an accelerometer. He does not need to compare his motion against some other reference frame to find out if he is accelerating. Newton was well aware of this fact, which led him to postulate the existence of 'absolute' space. Therefore, it appears that an accelerated test mass does *not* require another reference frame to gauge motion, and therefore acceleration has a special status in physics. However, it will be shown through the new quantum principle of inertia (discussed later), acceleration also has a special hidden reference frame that was unknown to both Einstein and Newton when they formulated their famous theories of motion. The reference frame in question here is the state of accelerated motion of the test mass with respect to the virtual particles of the quantum vacuum. However, it is not the velocity of the particles that sets up this abstract reference frame, it is the net statistical average **acceleration** of the virtual particles of the quantum vacuum near the test mass that forms the absolute reference frame. These concepts affect the meaning of inertial mass. Therefore, we elaborate on the meaning of inertial mass in the next section.

## 10. SPECIAL RELATIVITY - INERTIAL MASS AND INERTIAL FORCE

**"In contrast to the Newtonian conception, it is easy to show that in relativity the quantity force, is not codirectional with the acceleration it produces … It is also easy to show that these force components have no simple transformation properties ...."**

**- M. Hammer**

Quantum Inertia (QI) provides (section 11) a new understanding of Newtonian momentum. We will show that it is only *inertial force* (and forces in general) that is truly a *fundamental* concept of nature, not momentum or conservation of momentum. The Newtonian momentum, which is defined by 'mv', is simply a bookkeeping value used to keep track of the inertial mass 'm' (defined as F/A) in the state of constant velocity motion 'v' *with respect to another mass* that it might collide with at some future time. In this way, momentum is a relative quantity. Momentum simply represents information (with respect to some other mass) about what will happen in later (possible) force reactions. This fits in with the fact that *inertial* mass cannot be measured for constant velocity mass in motion (in outer space for example, away from all other masses) without introducing some **test** acceleration. If a mass is moving at a constant velocity, there are **no** forces present from the vacuum. Furthermore, since momentum involves velocity, it requires some other inertial reference frame in order to gauge the velocity 'v'. The higher the velocity that a mass 'm' achieves, the greater will be the subsequent deceleration (and therefore the greater the subsequent inertial force present) during a later collision (when it meets with some another object). If the velocity doubles with respect to a wall ahead, for example, then the deceleration doubles in a later impact. Before doubling the velocity, the acceleration $a_0 = (v_0 - 0)/t$; and after doubling, $a = (2v_0 - 0)/t = 2a_0$. Therefore we find that $f = 2f_0$, the force required from the wall (assuming the time of collision is the same). Similarly, if the mass is doubled, the force required from the wall doubles, or $f=2f_0$. Recall that inertial force comes from the *opposition of the quantum vacuum to the acceleration of mass* (or deceleration as in this case). Similarly, the kinetic energy '$1/2mv^2$' of a mass moving at a constant relative velocity 'v', it is also a bookkeeping parameter (defined as the product of <u>force</u> and the <u>time</u> that a force is applied). This quantity keeps track of the subsequent energy reactions that a mass will have when later accelerations (or decelerations) occur with respect to some other mass. It is important to remember that it is the *quantum vacuum force* that acts against an inertial mass to oppose any change in its velocity that is truly fundamental.

We therefore conclude that according to principles of QI theory, the inertial force is <u>absolute</u>. We have also seen that acceleration **can** be considered <u>absolute.</u> By this we mean that it is only the acceleration 'a' of a mass 'm' with respect to the net statistical average acceleration of the virtual particles of the quantum vacuum that accounts for inertial force. Therefore, we conclude that inertial mass can also be considered to be <u>absolute,</u> and follows the simple Newtonian relationship 'M=F/A'. Since inertial force, acceleration, and mass can all be considered to be absolute in this framework, we must closely reexamine the principles of special relativity in regards to the variation of inertial mass with the relative velocity of another inertial frame. Relativity is based on the premise that all constant velocity motion is relative, and also on the postulate of the constancy of light velocity. According to special relativity (which restricts itself to frames of constant velocity, called inertial frames), the inertial mass 'm' is relative, and varies with the relative velocity 'v' with respect to a constant velocity observer, in accordance with the following formula: $m = m_0 / (1-v^2/c^2)^{1/2}$. Here m is defined as the inertial mass measured in the other frame with velocity v, and $m_0$ is defined as the rest mass (inertial mass measured in the same frame as the mass) and 'c' is the velocity of light. It appears on the surface that QI and special relativity are not compatible in regards to the meaning of inertial mass. From the point of view of quantum inertia, Einstein's definition of inertial mass cannot be *fundamentally* correct, because it is not related to the quantum vacuum process described above for inertia. This is because we cannot associate the relative velocity 'v' directly to any quantum vacuum process. Recall that it is only the acceleration 'a' of a mass 'm', with respect to the net statistical average acceleration vectors of the virtual particles of the quantum vacuum that is the source of inertial mass.

Most special relativistic textbook accounts of inertial force and mass are based on the so-called 'conservation of momentum approach' (ref. 20). The conservation of momentum is assumed to be a fundamental aspect of nature. In order for momentum to be conserved with respect to all constant velocity reference frames, the mass must vary. To see this, recall that momentum is defined as mass times velocity, or 'mv', and that the momentum is important in a collision only because it provides bookkeeping of the mass and relative velocity. The *relative* velocity between the two colliding masses will determine the amount of deceleration in the impact as follows: $a=(v_f - v_i)/\Delta t$, were $v_f$ is the final velocity, and $v_i$ the initial velocity. Also, the mass is important because the subsequent force (and therefore energy $E = F \Delta t$) is determined by $F= m\ a$ through the quantum vacuum process described above. The more mass particles contained in a mass, the greater the resistance to the acceleration of the mass. Therefore, the product of mass and velocity is an indicator of the amount of *future* energy to be expected in a collision (or interaction) of the two masses. The total incoming momentum is defined as the momentum of the in-going masses ($m_1v_1 + m_2v_2$), the total out-going momentum is ($m_1v_1' + m_2v_2'$). Here the two masses $m_1$ and $m_2$ are moving at velocities $v_1$ and $v_2$ before the collision, with respect to an observer, and velocity $v_1'$ and $v_2'$ after the collision. In Newtonian mechanics, the total momentum is conserved for any observer in a constant velocity reference frame. Therefore, ($m_1v_1 + m_2v_2$) = ($m_1v_1' + m_2v_2'$), even though different observers in general will disagree with each of the relative velocities of a pair of masses that are colliding. This is what we mean by conservation of momentum. In special relativity, if we do not modify the definition of inertial mass, we would find that different observers in different constant velocity frames *disagree* on the conservation of momentum for colliding masses. However, it can be shown (ref. 20) that if the mass of an object 'm' (from the point of view of an observer in constant velocity motion 'v' with respect to a mass $m_0$, measured by an observer at relative rest) is redefined as follows:

$m = m_0 / (1-v^2/c^2)^{1/2}$ , then the *total* momentum of the collision remains conserved as in Newtonian mechanics.

How does special relativity treat the definition of inertial mass and inertial force? Since Einstein was aware that acceleration is not invariant in different inertial frames, he knew that Newton's law had to be modified.

To quote A.P. French (ref. 21):

*"... the discovery and specification of laws of force is a central concern of physics. It is certainly important, therefore, to know how to transform forces and equations of motion so as to give a description of them from the point of view of different inertial frames. Since in special relativity the acceleration is not invariant, we know that we cannot enjoy the simplicity of Newtonain mechanics, but we can certainly arrive at some useful and meaningful statements.*

*The starting point, which we indeed made use of in the initial stages of our approach to relativity is Newton's law in the form*

$$\mathbf{F} = d\mathbf{p}/dt = d(m\mathbf{v})/dt \qquad \text{where } m = m_0 (1-v^2/c^2)^{-1/2}$$

*We take this as a definition of $\mathbf{F}$. It is a natural extension (and simplest extension) of the non-relativistic result. It is not a statement that can be independently proved."*

French goes on to analyze the consequences of this relativistic force definition for components parallel and transverse with respect to the direction of acceleration:

*"... for the case where $F_{0x}$ is applied parallel to $\mathbf{v}$, causing an acceleration $a_{0x}$ ... $F_x = F_{0x}$ This is a striking result. Despite the change of the measures of mass and acceleration in the two frames, the measure of the x component of force remains the same.*

*When we make a similar calculation for the transverse force, we find that this invariance does not hold. ... i.e.* $\qquad F_y = 1/(1-v^2/c^2)^{1/2} \; F_{0y}$

*In the above results one can discern the feature that in general force and acceleration are not parallel vectors. ... Only in the instantaneous rest frame of a body can one guarantee that $\mathbf{F}$, as defined by the time derivative of momentum, is in the same direction as the acceleration."*

Einstein had to modify Newton's inertial law during his program to revise all physics in order to be relativistic, and was not aware of the existence of the quantum vacuum at that time. When Einstein considered this law, he found that in addition to incorporating his new relative mass definition formula above, he had to contend with relative accelerated motion. Contrary to popular belief, special relativity *does* address the problem of accelerated motion, which can be measured by any observer in an inertial reference frame. Therefore, in order to allow different observers in different states of constant velocity motion to measure inertial forces, Newton's law of motion must be changed. Since space and time are involved in measuring acceleration relative to an observer, and therefore acceleration must also be relative.

As we have seen in our QI analysis, ***inertial mass is absolute***. Furthermore, there exists absolute acceleration of a mass, which is defined as the state of acceleration of the matter particles making up that mass, with respect to the (net statistical) average acceleration of the virtual particles of the quantum vacuum. (Note: Since virtual particles interact with each other, not all the individual accelerations of the virtual particles with respect to the mass will be the same, hence the statistical nature of this statement). What about the applied force? As the force is applied, an acceleration results which causes the velocity of the mass to increase. What if the velocity of the mass approaches light speed with respect to the applied force? Is the force still as effective in further increasing the velocity of the mass?

## 10.1   RELATIVISTIC MASS VARIATION FROM PARTICLE EXCHANGES

Classical physics is based on the assumption that forces between two bodies can act upon each other instantaneously through direct contact. Furthermore, the resulting action is independent of the relative velocity of the two bodies from which the force acts. When Einstein proposed special relativity, he abolished all action-at-a-distance including forces acting instantaneously. However, forces were still treated by Einstein within the framework of classical Newtonian physics. In his program to make classical physics relativistic, he accepted Newton's law of inertial force without modification (in fact, the law F=MA was postulated as still being correct). Modern physics now treats **all** forces as a quantum particle exchange process. Note: The gravitational force is a special case where two exchange particles are involved (in EMQG, ref. 1, there are two different force exchange particles involved in gravity, simultaneously: the photon and the graviton particle). As an example, consider the electric force exchange process, which involves the photon particle as described by Quantum Electrodynamics (QED). Here the charged particles (electrons, positrons) act upon each other through the exchange of force particles, which are photons. In the language of computer science, electromagnetic forces can be viewed as being 'digital'. What appears to be a smooth force variation is really the result of countless numbers of photon exchanges, each one contributing a 'quanta' of electromagnetic force.

To see how the exchange process works for electromagnetic forces, we will examine the classical Coulomb force law in the rest frame of two stationary charges. The electric force from the two charged particles decrease with the inverse square of their separation distance (the inverse square law: $F = kq_1q_2/r^2$, where k is a constant, $q_1$ and $q_2$ are the charges, and r is the distance of separation). QED accounts for the inverse square law by the existence of an exchange of photons between the two electrically charged particles. The number of photons emitted by a given charge (per unit of CA time) is fixed and is called the charge of the particle. Thus, if the charge doubles, the force doubles because twice as many photons are exchanged during the force interaction. This force interaction process causes the affected particles to accelerate either towards or away from each other depending on whether the charge is positive or negative (different charges transmit photons with a slightly different wave functions). It is interesting to note that certain cellular automata patterns exhibit behaviors like charge. For example, in the famous 2D geometric CA called Conway's game of life there exists a class of CA patterns called 'guns', which constantly emit a steady stream of 'glider' patterns indefinitely. This CA emission process is constant without any degradation of the original gun pattern. This resembles the charge property possessed by electrons, where photons are constantly emitted without any change of state of the electron.

The strength of the electromagnetic force depends on the quantity of the electric charge, and also depends on the distance of separation between the charges in the following way: each charge sends and receives photons from every direction. But, the number of photons per unit area, emitted or received, decreases by the factor $1/4\pi r^2$ (the surface area of a sphere) at a distance 'r', because the photon emission process take place in all directions. Thus, if the distance doubles, the number of photons exchanged decreases by a factor of four. This process can be easily visualized on a 3D geometric CA. Imagine that an electron is at the center of a sphere and sends out virtual photons in all directions. Imagine that a second electron somewhere on the surface of a sphere at a distance 'r' from the emitter, absorbing some of the exchange photons. The absorption of the exchange photons causes an outward acceleration, and thus a repulsive force. If the charge is doubled on the electron, there is twice as many photons appearing at the surface of the sphere, and twice the force acting on the electron. Thus, this accounts for the linear product of charge terms in the inverse square law. In QED, photons do not interact with photons (by a force exchange interaction). As a result, in-going and out-going photons do not affect each other during the exchange process.

The details of the particle exchange process for gravity is covered in EMQG (ref. 1). For now we are interested in the consequences of the force exchange process for special relativity, where the exchange particle has a <u>finite, and fixed velocity of propagation</u> (the speed of light, with the exception of the weak nuclear force where some bosons carry mass). To our knowledge, no one has examined the consequences of particle exchanges from the point of view of forces acting on each other in different inertial frames, where exchange particle propagates at the speed of light. At the time that Einstein developed special relativity, the force exchange process was unknown. The basic idea we want to develop here is that the quantity of force transmitted between two objects very much depends on the received <u>flux rate</u> of the exchange particles. In other words, the number of particles exchanged per unit of <u>time</u> represents the magnitude of the force transmitted between the particles. For example, imagine that there are two charged particles at relative rest in an inertial reference frame. There are a fixed number of particles exchanged per second at a separation distance 'd'. Now imagine that particle B is moving away at a slow constant relative velocity 'v' with respect to particle A. If the relative velocity v<<c the exchange process appears almost the same as when the two particles are at rest. This is because the velocity of light is very high when compared to 'v', and the flux rate is unaffected. Now imagine that the relative recession velocity v -> c, which is comparable to the velocity of the exchange particle. Does the received flux rate of particle B get altered from the perspective of particle B's frame? The answer is yes, and this follows from another result of special relativity: the Lorentz Time Dilation!

It is clear from Lorentz time dilation that the timing of the exchange particle will be altered when there is a very high relative velocity away from the source. Recall the Lorentz time dilation formula of special relativity: $t = t_0 / (1 - v^2/c^2)^{1/2}$, which states that the timing of events varies with relative velocity 'v'. If the timing of the exchange particles is altered, then the flux rate is altered as well, since flux has units of numbers of particles per unit time.

Now assume that particle A emits a flux of $\Phi_a$ particles per second, as seen by an observer in particle A's rest frame. When the force exchange particles are transmitted to particle B, particle B sees the flux rate decrease because of time dilation. Therefore, we find that particle B receives a smaller quantity of exchange particles per second $\Phi_b$ then when the particles are at relative rest. Thus, particle A acts like it transmits a smaller flux rate $\Phi_b$, such that $\Phi_b = \Phi_a (1 - v^2/c^2)^{1/2}$. Since the force due to the particle exchange is directly proportional to the flux of particles exchanged, we can therefore write:

$$F = F_0 (1 - v^2/c^2)^{1/2} \tag{10.1}$$

where is $F_0$ is the magnitude of the force when particle A and B are at relative rest, and F is the resulting smaller force acting between particle A and B when the receding relative velocity is 'v'. Thus, we can conclude that when a force acts to cause an object to recede away from the source of the force, the force <u>reduces</u> in strength. With a similar line of reasoning, we find that the force increases in strength when a force acts to cause an object to move towards the source of the force.

We are now in a position to see the apparent relationship between the inertial mass and velocity. Since all forces are due to particle exchanges, we can use the method developed above to study the forces acting between to inertial frames. First, at relative rest where v=0, we have $F = F_0$. The rest mass '$m_0$' is defined by Newton's law: $F_0 = m_0 a$, where 'a' is a test acceleration that is introduced to measure the inertial rest mass. Now, assume that there is a relative velocity 'v' between the applied force and the mass 'm', which causes the mass to recede. Therefore, we can write:

$$F = F_0 (1 - v^2/c^2)^{1/2} = m a \tag{10.2}$$

where the force is reduced in magnitude for the reasons discussed above, and the mass 'm' is considered absolute (or $m = m_0$, as in Newtonian Mechanics). We believe that equation 10.2 represents the actual physics of the force interaction. However, if one takes the position that the force does not vary with velocity, but that the mass is what actually varies, then the above equation can be interpreted as:

$F = F_0 = m a = m_0 (1 - v^2/c^2)^{-1/2} a$ , and  $m = m_0 (1 - v^2/c^2)^{-1/2}$ as given by Einstein.

So we see that we are in a situation where it is experimentally impossible to distinguish between the following two approaches: inertial mass variation with high velocity (Einstein) versus the force variation with high velocity. What velocities can a mass achieve through the application of an accelerating force? According to our analysis above, the answer is that the limiting speed is the speed of the exchange particles, or light velocity. At this limit, the accelerating force effectively becomes zero!

It is, however, convenient to associate the variation of force with an increase in relativistic mass as Einstein proposed, for two important reasons. First this restores the conservation of total momentum in collisions for all inertial observers (in fact, this is how Einstein derived his famous mass-velocity relationship). Secondly, if a mass is accelerated to the relativistic velocity 'v' with respect to observer 'A' by some given force, and this force is removed, there will be no way to determine the subsequent energy release when a collision occurs later. In other words, when this mass collides with another object, a rapid deceleration occurs with a large release of energy (which is force multiplied by time). This energy release is greater then what can be expected from Newton's laws. In fact, the large energy release is due to the effective increase in the force during the collision due to increased numbers of force exchange particles acting to reduce the speed of the colliding mass.

The force $F = F_0 (1 - v^2/c^2)^{1/2}$ tends to zero as the velocity $v \rightarrow c$. This means that any force becomes totally ineffective as the mass is accelerated to light velocity with respect to the source. As we have seen, this is attributed to the force resulting from exchange particles, which become totally ineffective in propagating from the source to the receiver, as the velocity of the receiver with respect to the source approach the velocity of the exchange particle. In order to clarify these ideas, we will analyze an actual experiment that was performed to confirm relativistic mass increase effect.

10.2   ANALYSIS OF 'ULTIMATE SPEED' EXPERIMENT FOR ELECTRONS

The textbook titled 'Special Relativity' by A.P. French (ref. 21) describes an actual experimental test of relativistic mass and energy, which was performed for a film called 'The Ultimate Speed'. This experiment is designed to measure special relativistic effects such as mass increase, momentum, and energy of an electron accelerated to relativistic velocities. The experiment consists of an electron gun, a linear accelerator, an oscilloscope to measure the electron time of flight, and an aluminum disk to stop the electrons and signal the arrival. A direct calorimetric measurement of the electrons is made in order to compare the energy transferred by the electrons during collision with the target. We will give our interpretation of this experiment in terms of the quantum vacuum interactions involved for mass, and the variation of accelerating force with relative velocity discussed above.

The electrons are emitted from the electron gun and accelerated through an electrical potential in the laboratory frame. We assume that the electrical force acts in a direction that is parallel to the direction of motion. An electrical force is required to accelerate the electrons in order to overcome the resistance offered by the charged virtual particles of the quantum vacuum, which is simply the electron's inertia. (Note: This is the definition of inertial mass in EQMG, where each charged particle inside the electron contributes elementary quanta of inertia). This resistance to acceleration is the <u>absolute inertial mass</u> (or rest mass in relativity), where absolute mass is the *actual* resistance to acceleration originating from the vacuum. As the electrons are accelerated, photon particle exchanges occur between the charged electrons and the accelerating plates maintained at a high electrical potential. When the velocity of the electrons with respect to the accelerating plates in the laboratory frame is much smaller then the speed of light, the force exchange process is unaffected. In other words the accelerating electrical force is just as effective in accelerating the electrons as if the electrons are at rest, with the force being qV (q is the electron charge, and V is the accelerating potential. As the velocity of the electron approaches light velocity however, the timing of arrival of the exchange particles is delayed in the electron frame as described previously. The force on the electron is reduced to:

$F = F_0 (1 - v^2/c^2)^{1/2}$ , where $F_0$ is the magnitude of the force at low velocities.

Table 1-1 of ref. 21 tabulates the time of flight, the kinetic energy, and the velocity of the electron. It was observed that the velocity of the electron increased more slowly as the electron achieved relativistic speeds, in spite of the large forces applied. To quote A. P. French:

 "… *the kinetic energy is raised by a factor of 30, so one might have looked for a factor of 5.5 in the speed (according to Newtonian Mechanics). Instead, there is an increase of only 15% (in the speed). The increase of v between 1.5 and 4.5 MeV is barely detectable within the accuracy of the experiment. One might therefore question whether the electrons are in fact being given the energy calculated from the value of qV …*"

In order to verify that the electrons actually have the energies calculated, the experimenter makes a *direct* calorimetric measurement of the collision energy of the electrons with the aluminum target plate. They conclude that the energy given to the electron (force multiplied by time) is definitely carried by the electrons during this collision, through their heat dissipation measurement.

What actually happens in the experiment however, is that the force becomes less and less effective in increasing the velocity of the electrons during the relativistic phase of the electron flight, in the reference frame of the accelerating electrons. In other words, the electrons achieve little increase in speed (a = f / m, and since the force 'f' decreases, the acceleration 'a' decreases, m remaining constant) for the force applied. Where does the extra energy come from during the collision (as compared to Newtonian mechanics)? The collision of the electrons with the aluminum target is really an extremely rapid deceleration, and therefore involves the quantum vacuum. The electrons are rapidly decelerated by the aluminum target, which creates a relative acceleration between the charged electrons and the background, charged virtual particles of the quantum vacuum. However, in the reference frame of the relativistic electrons, the number of exchange particles per second increases (as compared to the slow velocity phase of the electrons), creating a larger force operating against the target. One can view the exchange particles as being 'red shifted' in terms of frequency (particles per second) during the acceleration phase of the electron trajectory giving a decrease in force, and 'blue-shifted' during the deceleration phase in collision giving an increase in force (and thus energy).

To summarize, **inertial mass is absolute.** The applied force that is used to accelerate a mass is reduced at relativistic speeds, and depends on the relative velocity between the applied force and the accelerated mass. The absolute acceleration of a mass is defined as the state of acceleration of the matter particles making up that mass with respect to the (net statistical) average acceleration of the surrounding virtual particles of the quantum vacuum.

10.3   EQUIVALENCE OF MASS AND ENERGY: $E = M c^2$

One of the most important results of special relativity is the equivalence between mass and energy. This is represented in perhaps the most famous formula in all of physics: $E=Mc^2$. This formula implies that photons carry mass, since they carry energy. Photons are capable of transferring energy from one location to another, as by solar photons for example. Do photons really have mass?

You might think that if a particle has energy, it automatically has mass; and if a particle has mass, then it must emit or absorb gravitons. This reasoning is based of course, on $E=mc^2$. Einstein derived this formula from his famous light-box thought experiment (ref. 20). In his thought experiment, a photon is emitted from a box, causing the box to recoil and thus to change momentum. In quantum field theory this momentum change is traceable to a fundamental QED vertex, where a electron (in an atom in the box) emits a photon, and recoils with a momentum equivalent to the photon's momentum '$m_p c$". Therefore, we can conclude that the photon behaves as if it has an effective inertial mass '$m_p$' given by: $m_p = E/c^2$ in Einstein's light box. For simplicity, lets consider a photon that is absorbed by a charged particle like an electron at rest. The photon carries energy and is thus able to do work. When the photon is absorbed by the electron with mass '$m_e$', the electron recoils, because there is a definite momentum transfer to the electron given by $m_e v$, where v is the recoil velocity. The electron momentum gained is equivalent to the effective photon momentum lost by the photon $m_p c$. In other words, the electron momentum '$m_e v_e$' received from the photon when the photon is absorbed is equivalent to the momentum of the photon '$m_p c$', where $m_p$ is the effective photon mass. If this electron later collides with another particle, the same momentum is transferred. The rest mass of the photon is defined as zero. Thus, the effective photon mass is a measurable inertial mass.

**Note**: the recoil of the light box is a backward acceleration of the box, which works against the virtual particles of the quantum vacuum. Thus, when one claims that a photon has a real mass, we are really referring to the photon's ability to impart momentum. This momentum can later do work in a quantum vacuum inertial process.

Einstein's derivation of E=mc$^2$ was unnecessarily complex (ref. 20) because of his reluctance to utilize results from quantum theory. Although he was one of the founders of the (old) quantum theory, he remained skeptical about the validity of the theory throughout his whole career. We treat the energy-mass equivalence as a ***purely*** quantum process, and not as a result of special relativity. Although Einstein derived this law when he developed special relativity, it can be derived purely from quantum theory. As we hinted, the ability of a photon to transfer momentum (and thus carry energy) can be traced to a QED vertex, where a packet of momentum is transferred from the photon to an electron. Let us assume that the effective mass of the photon is $m_c$. Furthermore, the photon has a velocity c, momentum p, energy E, a wavelength λ, and a frequency ν. Therefore, by using the properties of the photon below (where h is plank's constant):

P=mc   (Momentum defined by classical physics)                                      (10.31)

c=νλ   (from the classical definition of frequency and wavelength)         (10.32)

E=hν   (from Plank's quantum energy-frequency law)                               (10.33)

λ=h/p   (from Quantum DeBroglie wavelength-momentum law)            (10.34)

Therefore, c/ν = h/p = h/(mc)  (using 10.32, 10.34, and 10.31).

Finally, c/(E/h) = h/(mc) (using 10.33), or E=mc$^2$.

Thus, a very simple derivation of the energy-mass relationship is possible from the basic results of quantum mechanics.

## 11.    THE QUANTUM THEORY OF INERTIA

**"Under the hypothesis that ordinary matter is ultimately made of subelementary constitutive primary charged entities or 'partons' bound in the manner of traditional elementary Plank oscillators, it is shown that a heretofore uninvestigated Lorentz force (specifically, the magnetic component of the Lorentz force) arises in any accelerated reference frame from the interaction of the partons with the vacuum electromagnetic zero-point-field (ZPF). ... The Lorentz force, though originating at the subelementary parton level, appears to produce an opposition to the acceleration of material objects at a macroscopic level having the correct characteristics to account for the property of inertia."**

<div style="text-align: right">**- B. Haisch, A. Rueda, H. E. Puthoff**</div>

According to CA theory, there must be a localized explanation for all global phenomena such as acceleration and gravity. Inertia and gravity should originate from the small-scale particle interactions such that a global law emerges from the activity. Recall that CA theory is based on the local rules for the local cellular neighborhood, and these rules are repeated on a vast scale for all the cells in the universe. Many of our existing physical theories are general, global principles or general observations of nature. Both gravity and inertia have only been described successfully by "classical theories", applicable on global scales. In EQMG, both inertia and gravity have a detailed, particle level explanation based on the local "conditions" at the neighborhood of a given matter particle, and is thus manifestly compatible with the philosophy of a cellular automata theory and the principle of locality in special relativity.

In a recent theory (ref. 5) proposed by Haisch, Rueda, and Puthoff (known here as the HRP Theory of Inertia), it was shown that inertia comes from the buzz of activity of the virtual particles that fills even a perfect vacuum. It is this ever-present sea of energy that resists the acceleration of mass, and so creates inertia. Thus, they have found the low-level quantum description of inertia that is manifestly compatible with CA theory. Inertia is now described as being purely the result of quantum particle interactions. Haisch, Rueda, and Puthoff have come up with a new version of Newton's second law: F=MA. As in Newton's theory, their expression has 'F' for force on the left-hand side and 'A' for acceleration on the right. But in the place of 'M', there is a complex mathematical expression tying inertia to the properties of the vacuum. They found that the fluctuations in the vacuum interacting with the charge particles of matter in an accelerating mass give rise to a magnetic field, and this in turn, creates an opposing force to the motion. Thus, electromagnetic forces (or photon exchanges) is ultimately responsible for the force of inertia! The more massive an object, the more 'partons' it contains; and the more partons a mass contains means more individual (small) electromagnetic forces from the vacuum are present and the stronger the reluctance to undergo acceleration. But, when a mass is moving at a **constant** velocity, inertia disappears, and there is no resistance to motion in any direction as required by special relativity.

In their theory, inertia is caused by the magnetic component of the Lorentz force which arises between what the author's call 'parton' particles in an accelerated reference frame interacting with the background vacuum electromagnetic zero-point-field (ZPF). The author's use the old fashion term originated by Feynman called the 'parton', which referred to the elementary constituents of the nuclear particles such as protons and neutrons. It is now known that Feynman's partons are quarks, and that the proton and neutron each contain three quarks of two types: called the 'up' and 'down' quarks.

According to EMQG (ref. 1) quantum inertia is also deeply connected with the subject of quantum gravity. EMQG explains why the inertial mass and gravitational mass are identical in accordance with the weak equivalence principle. The weak equivalence principle translates to the simple fact that the mass (m) that measures the ability of an object to produce (or react to) a gravitational field ($F=GMm/r^2$) is the same as the inertial mass value that appears in Newton's F=ma. In EMQG, this is not a chance coincidence, or a given fact of nature, which is assumed to exist, *a prior*. Instead, equivalence follows from a deeper process occurring inside a gravitational mass due to interactions with the quantum vacuum, which are *very similar* in nature to the interactions involved in inertial mass undergoing acceleration.

Since this new quantum theory of the inertia has still not been fully developed or confirmed yet, we raise QI to the level of a postulate. The virtual particles of the quantum vacuum can be considered to be a kind of absolute reference frame for accelerated motion only. This frame is simply represented as the resultant acceleration vector given by the sum of all the acceleration vectors of the virtual particles of the quantum vacuum in the immediate neighborhood of a given charged masseon particle in the accelerated mass. This quantum vacuum reference frame gauges absolute acceleration. We do not need to measure our motion with respect to this frame in order to confirm that a mass is accelerated, we simply need to measure if an inertial force is present. We will see that this new, local quantum vacuum reference frame resolves Mach's paradox in regards to what reference frame nature uses to gauge accelerated mass.

## 12. APPLICATIONS OF QUANTUM INERTIA: MACH'S PRINCIPLE

*"... it does not matter if we think of the earth as turning round on its axis, or at rest while the fixed stars revolve around it ... The law of inertia must be so conceived that exactly the same thing results from the second supposition as from the first."*

<div style="text-align: right">*E. Mach*</div>

Ernst Mach (ca 1883) proposed that the inertial mass of a body does not have any meaning in the absence of the rest of the matter in the universe. In other words, acceleration requires some other reference frame in order to determine accelerated motion. Thus, it seemed to Mach that the only reference frame possible was that of the average motion of all the other masses in the universe. This implied to Mach that the acceleration of an object must somehow be dependent on the sum total of all the matter in the universe. To Mach, if all the matter in the universe were removed, the acceleration, and thus the force of inertia would completely disappear since no reference frame is available to determine the actual acceleration.

A spinning elastic sphere bulges at the equator due to the centrifugal force. The question that Mach asked was how does the sphere 'know' that it is spinning, and therefore must bulge. If all the matter in the universe was removed, how can we be sure that it really rotates? Therefore, how would the sphere 'know' that it must bulge or not? Newton's answer would have been that the sphere somehow felt the action of Newtonian absolute space. Mach believed that the sphere somehow 'feels' the action of all the cosmic masses rotating around it. To Mach, centrifugal forces are somehow gravitational in the sense that it is the action of mass on mass. To Newton, the centrifugal force is due to the rotation of the sphere with respect to absolute space. To what extent that Einstein's general theory of relativity incorporates Mach's ideas is still a matter of debate. We (through the quantum inertia principle) take a similar view as Newton, where Newton's absolute space is replaced by the virtual particles of the vacuum. Mach was never unable to develop a full theory of inertia based on his idea of mass affecting mass.

Mach's ideas on inertia are summarized as follows:

(1) A particle's inertia is due to some (unknown) interaction of that particle with all the other masses in the universe.
(2) The local standards of non-acceleration are determined by some average value of the motion of all the masses in the universe.
(3) All that matters in mechanics is the relative motion of all the masses.

Quantum inertia theory fully resolves Mach's paradox by introducing a new universal reference frame for gauging acceleration: the net statistical average acceleration vector of the virtual particles of the quantum vacuum with respect to the accelerated mass. In other words, the cause of inertia is the interaction of each and every particle with the quantum vacuum. Inertial force actually *originates* in this way. It turns out that the distant stars do affect the local state of acceleration of our vacuum here through the long-range gravitation force. Thus, our local inertial frame is slightly affected by all the masses in the distant universe. However, in our solar system the local gravitational bodies swamp out this effect. (This long-range gravitational force is transmitted to us by the graviton particles that originate in all the matter in the universe, which will distort our local net statistical average acceleration vector of the quantum virtual particles in our vacuum with respect to the average mass distribution). Thus, it seems that Mach was correct in saying that acceleration here depends somehow on the distribution of the distant stars (masses) in the universe, but the effect he predicted is minute.

## 12.1   MORE APPLICATIONS OF QI: NEWTON'S LAWS OF MOTION

We are now in a position to understand the quantum nature of Newton's classical laws of motion. According to the standard textbooks of physics (ref. 19) Newton's three laws of laws of motion are:

**1. An object at rest will remain at rest and an object in motion will continue in motion with a constant velocity unless it experiences a net external force.**

**2. The acceleration of an object is directly proportional to the resultant force acting on it and inversely proportional to its mass. Mathematically: $\Sigma F = ma$, where 'F' and 'a' are the vectors of each of the forces and accelerations.**

**3. If two bodies interact, the force exerted on body 1 by body 2 is equal to and opposite the force exerted on body 2 by body 1. Mathematically: $F_{12} = -F_{21}$.**

Newton's first law explains what happens to a mass when the resultant of all external forces on it is zero. Newton's second law explains what happens to a mass when there is a nonzero resultant force acting on it. Newton's third law tells us that forces always come in pairs. In other words, a single isolated force cannot exist. The force that body 1 exerts on body 2 is called the action force, and the force of body 2 on body 1 is called the reaction force.

Newton's first two laws are the direct consequence of the (electromagnetic) force interaction of the (charged) elementary particles of the mass interacting with the (charged) virtual particles of the quantum vacuum. Newton's third law of motion is the direct consequence of the fact that all forces are the end result of a boson particle exchange process.

Newton's First Law of Motion:

The first law is a trivial result, which follows directly from the quantum principle of inertia (postulate #3). First a mass is at relative rest with respect to an observer in deep space. If no external forces act on the mass, the (charged) elementary particles that make up the mass maintain a *net acceleration* of zero with respect to the (charged) virtual particles of the quantum vacuum through the electromagnetic force exchange process. This means that no change in velocity is possible (zero acceleration) and the mass remains at rest. Secondly, a mass has some given constant velocity with respect to an observer in deep space. If no external forces act on the mass, the (charged) elementary particles that make up the mass also maintain a *net acceleration* of zero with respect to the (charged) virtual particles of the quantum vacuum through the electromagnetic force exchange process. Again, no change in velocity is possible (zero acceleration) and the mass remains at the same constant velocity.

Newton's Second Law of Motion:

The second law is the quantum theory of inertia discussed above. Basically the state of *relative* acceleration of the charged virtual particles of the quantum vacuum with respect to the charged particles of the mass is what is responsible for the inertial force. By this we mean that it is the tiny (electromagnetic) force contributed by each mass particle undergoing an acceleration 'A', with respect to the net statistical average of the virtual particles of the quantum vacuum, that results in the property of inertia possessed by all masses. The sum of all these tiny (electromagnetic) forces contributed from each charged particle of the mass (from the vacuum) is the source of the total inertial resistance force opposing accelerated motion in Newton's F=MA. Therefore, inertial mass 'M' of a mass simply represents the total resistance to acceleration of all the mass particles.

Newton's Third Law of Motion:

According to the boson force particle exchange paradigm (originated from QED) all forces (including gravity, as we shall see) result from particle exchanges. Therefore, the force that body 1 exerts on body 2 (called the action force), is the result of the emission of force exchange particles from (the charged particles that make up) body 1, which are readily absorbed by (the charged particles that make up) body 2, resulting in a force acting on body 2. Similarly, the force of body 2 on body 1 (called the reaction force), is the result of the absorption of force exchange particles that are originating from (the charged particles that make up) body 2, and received by (the charged particles that make up) body 1, resulting in a force acting on body 1. An important property of charge is the ability to readily emit <u>and</u> absorb boson force exchange particles. Therefore, body 1 is both an emitter and also an absorber of the force exchange particles. Similarly, body 2 is also both an emitter and an absorber of the force exchange particles. This is the reason that there is both an action and reaction force. For example, the contact forces (the mechanical forces that Newton was thinking of when he formulated this law) that results from a person pushing on a mass (and the reaction force from the mass pushing on the person) is really the exchange of photon particles from the charged electrons bound to the atoms of the person's hand and the charged electrons bound to the atoms of the mass on the quantum level. Therefore, on the quantum level there is really is no contact here. The hand gets very close to the mass, but does not actually touch. The electrons exchange photons among each other. The force exchange process works both directions in equal numbers, because all the electrons in the hand and in the mass are electrically charged and therefore the exchange process gives forces that are equal and opposite in both directions.

## 13. CONCLUSIONS

We have presented a new paradigm for physical reality, which restores a great unity to all physics. First, we have argued that our universe is a vast Cellular Automata (CA) simulation, the most massively parallel computer model known. This CA structure is a simple 3D geometric CA. All physical phenomena, including space, time, matter, and forces are the result of the interactions of *information* patterns, governed by the mathematical laws and the connectivity of the CA. Because of the way the CA functions, all the known global laws of the physics must result from the local mathematical law that governs each cell, and each cell contains the same mathematical law. We have seen that the CA structure automatically presents our universe with a speed limit for 'motion'. Quantum field theory requires that all forces are the result of the boson particle exchange process. This particle exchange paradigm fits naturally within CA theory, where the boson exchange represents the transfer of boson information patterns between (fermion) matter particles. All forces (gravity is no exception) originate from exchange processes as dictated by quantum field theory. The two basic postulates of special relativity were reviewed using CA theory. We found that both postulates can be ***derived*** from our CA model.

The first postulate of special relativity states that the velocity of light in a vacuum is constant and is equal for all observers in inertial frames (inertial frame is one in which Newton's law of inertia is obeyed). We have shown that ***any*** Cellular Automata model inherently has a maximum speed at which information can be transferred from cell to cell. We introduced a new postulate regarding the nature of photon motion on the CA. The photon information pattern is *only* capable of shifting one cell per clock period, and not at any other rate, and therefore exits at one speed with respect to the absolute cell locations. The value for the speed of light can then be derived simply as the ratio of (our) distance over (our) time for the information pattern transfer rate. The maximum information transfer velocity is: $V_P = L_P / T_P = 3 \times 10^8$ meter/sec = c. This is the Plank velocity, and has a value of one in absolute units.

*(NOTE: In EMQG this maximum speed actually represents the raw or 'low-level' light velocity, defined as the velocity of light in between encounters with virtual particles. The scattering of photons with the virtual particles of the quantum vacuum reduces the speed of the photons to the familiar observable light velocity through a process known as photon scattering with the virtual particles. This complicates the simple picture presented here. Refer to reference 1 for the details of this process).*

The second postulate of special relativity states that the laws of physics are equally valid in all inertial reference frames. This follows from CA theory by remembering that all cells and their corresponding rules in the cellular automata are absolutely identical everywhere. Motion itself is an illusion, and really represents information transfers from cell to cell. To assign meaning to motion in a CA, one must relate information pattern flows from one numeric pattern group with respect to another group (the actual cell locations are inaccessible to experiment). Therefore, motion requires reference frames. Unless you have access to the absolute location of the cells, all motion remains relative. In other words, there is *no* reference frame accessible by *experiment* that can be considered as the absolute reference frame for constant velocity motion. In Quantum Inertia, the virtual particles of the quantum vacuum still do not allow us to reveal our (constant velocity) motion between two inertial frames. However, this is not the case for an accelerated observer or for observers in gravitational fields, where the virtual particles of the quantum vacuum plays a major role as an absolute reference frame for accelerated motion only

We presented a new theory of Newtonian inertia called Quantum Inertia. It was based on the idea that inertial force is due to the tiny electromagnetic force interactions (where the details of this force have not been fully defined at this time) originating from each (charged) particle of real matter undergoing relative acceleration with respect to the virtual, electrically charged matter particles of the quantum vacuum. These tiny forces are the source of the total resistance force to accelerated motion in Newton's law 'F = MA', where the sum of each of the tiny forces equals the total inertial force. The exact details of these forces are not known, and hence Quantum Inertia remains a postulate. This new approach to classical inertia automatically resolves the problems and paradoxes of accelerated motion introduced in Mach's principle, by suggesting that the virtual particles of the quantum vacuum serve as Newton's universal reference frame (which he called absolute space) for accelerated motion only. Newton was correct in that it is the relative accelerated motion with respect to absolute space that somehow determines the inertia of a mass, but absolute space is totally useless in determining the *absolute* velocity of an object.

The relative accelerated motion of the virtual particles of the quantum vacuum with respect to the average motion of the real matter particles that are bound in a mass can be used as the reference frame to define absolute mass (which is equivalent to special relativistic rest mass). This is contrary to special relativity, and to the known mass-velocity formula of special relativity. However, we have shown that the quantity of force transmitted between two objects in different inertial reference frames depends on the flux rate of the exchange particles. In other words, the number of particles exchanged per unit of time represents the magnitude of the force transmitted between the particles. If the relative velocity $v \ll c$, then the exchange process appears almost the same as when the two particles are at rest. Now as the relative recession velocity $v \to c$, it is comparable to the velocity of the exchange particle, and this affects the received flux rate. From the Lorentz time dilation, the timing of the exchange particle is altered, as given by $t = t_0 / (1 - v^2/c^2)^{1/2}$. The timing of the exchange particles is altered, and therefore the flux rate is altered as well, since flux has units of numbers of particles per unit time. Therefore, the magnitude of the force is reduced by: $F = F_0 (1 - v^2/c^2)^{1/2}$; where is $F_0$ is the magnitude of the force when at relative rest. Thus, we have concluded that the force is actually *reduced* in strength, and not that the mass increases. The reduced force is less effective in accelerating a mass, which can be interpreted as a mass increase.

Minkowski 4D space-time is shown to be a consequence of the behavior of matter (elementary particles) and light (photons) on the Cellular Automata. At the tiniest quantum distance scales there exists a kind of secondary (quantized) absolute CA 3D space and separate absolute (quantized) time, as required by CA theory. This forms a very special universal reference frame on the CA, and this frame is completely hidden from direct experimentation. This is the 'CA absolute reference frame'. We can associate with this frame an *absolute space* and *absolute time*. Everyday objects occupies a specific volume of cells in CA space. These cell patterns are (most likely) shifting through cell space at some given rate. Therefore, there exists a kind of absolute space and absolute time scale, but these are hidden from view and cannot be accessed experimentally.

Absolute CA 3D space and time is represented by a rectangular 3D array of numbers or cells; $C(x,y,z)$. These cells change state after every new CA clock operation $\Delta t$. The array of numbers $C(x,y,z)$ is called CA space, which acts like the Newtonian version of Cartesian absolute space. There also exists a separate absolute time needed to evolve the numerical state of the CA. 3D CA space and time are not effected by any physical interactions, and are also not accessible through direct measurement. In CA space, the Plank distance ($1.6 \times 10^{-35}$ meters) scale roughly corresponds to the minimum event distance (or cell 'size'), and Plank time scale ($5.4 \times 10^{-44}$ sec) to the minimum time interval possible.

The Lorentz transformation follows mathematically from the two postulates of special relativity. We have carefully analyzed the measurement of light velocity with respect to two fictitious inertial observers who somehow have access to the measurement of absolute CA units of space and time. We found that in absolute units, the velocity of light is the same for both observers, but the measurements of space and time varies. This is precisely what happens to the two real observers with real measuring instruments. However, the relative 4D space-time is simply an illusion, based on the ignorance of the actual state of information by the real observer.

Thus, we found that special relativity follows directly from the CA model of the universe, a vast 3D Cellular Automata computer. In the process, we have discovered the true nature of light motion and the principle of relativity. In the process, we have revealed the hidden quantum processes behind classical Newtonian Inertia.

**FIGURE CAPTIONS:**

(1) **Figure #1 - Block Diagram of the 3D Geometric Cellular Automata**
(2) **Figure #2 - Simplified Model of the Motion of a Photon Information Pattern**
(3) **Figure #3 - Light Velocity Measurement in absolute CA Units of Space and Time**
(4) **Figure #4 - Definition of an Inertial Reference Frame**

In a 3D Geometric Cellular Automata the numeric content of Cell $C_{i,j,k}$ is determined by the numeric contents of each of the surrounding 26 neighbours (and possibly with it's own numeric content). On the next CA 'clock cycle', the contents of cell $C_{i,j,k}$ is determined by a unique Function (or algorithm) $F(i,j,k)$ such that: $C_{i,j,k} = F(C_{i+x,j+y,k+z})$ where x,y,z take on these values: -1,0,1. This same function is programed in every cell.

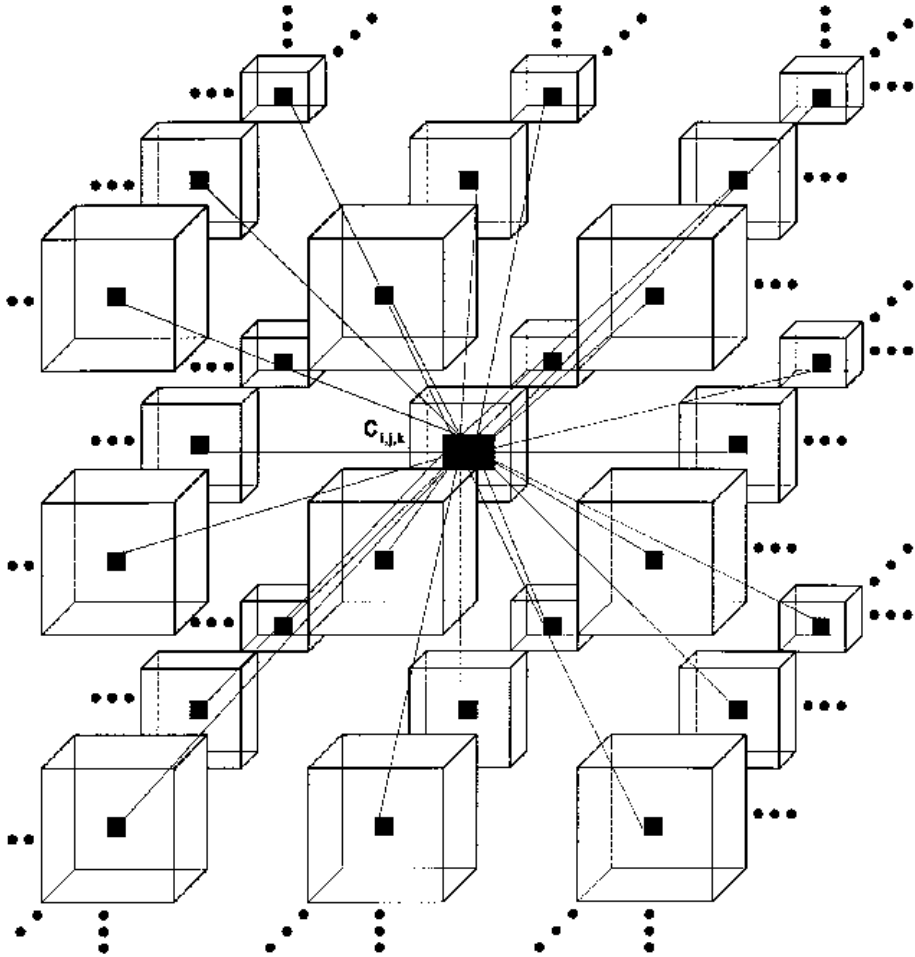

Figure # 1   - Block Diagram of the 3D Geometric Cellular Automata

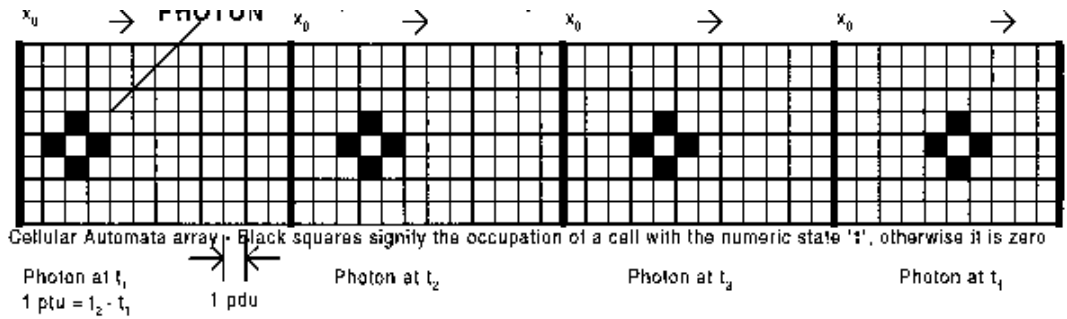

**Figure #2** - Simplified model of the motion of the photon information pattern on the CA.
The photon information pattern moves 1 plank unit to the right at every plank 'clock cycle'
(Note: The photon is actually an oscillating wavepattern (the wavefunction not shown in this simplified diagram)

**Absolute CA units:** 1 pdu is the shifting of information by 1 cell; 1 ptu is the time to shift 1 cell; 1 pvu = photon velocity

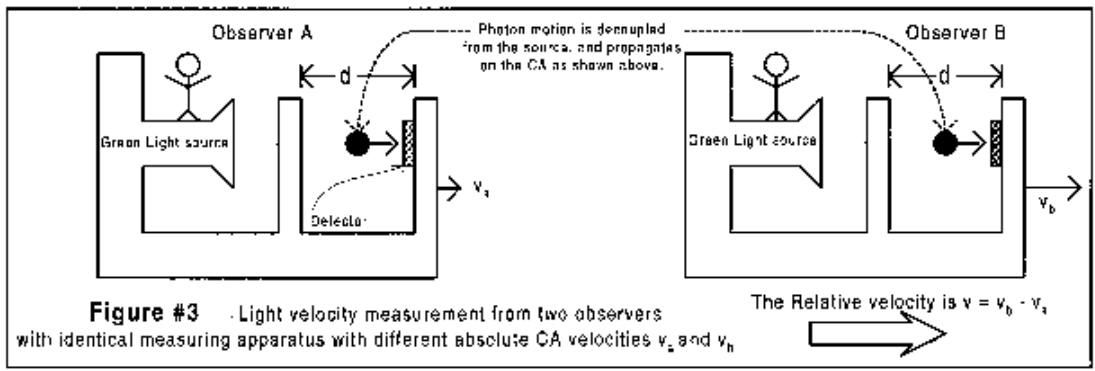

**Figure #3** - Light velocity measurement from two observers with identical measuring apparatus with different absolute CA velocities $v_a$ and $v_b$

The Relative velocity is $v = v_b - v_a$

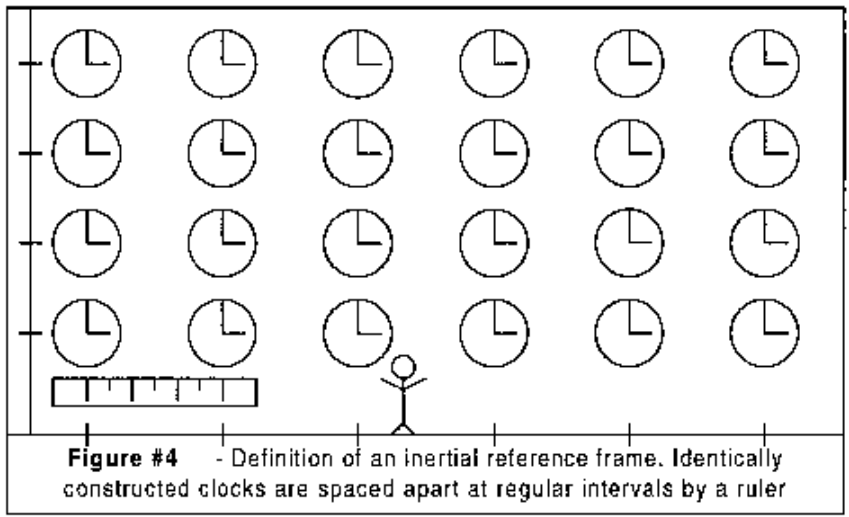

**Figure #4** - Definition of an inertial reference frame. Identically constructed clocks are spaced apart at regular intervals by a ruler